\documentclass[twocolumn, tighten]{aastex6}

\newcommand{\editbf}[1]{{#1}}

\usepackage{color}
\usepackage{amsmath}
\usepackage{graphicx}
\usepackage{natbib}
\usepackage{multirow}
\usepackage{booktabs}
\usepackage{tabularx}
\usepackage[normalem]{ulem}
\usepackage{array}

\definecolor{lightgray}{gray}{0.9}
\newcolumntype{L}{>{\centering\arraybackslash}m{3cm}}

\graphicspath{{figures/}}

\slugcomment{Accepted to The Astronomical Journal, June 2016}

\shorttitle{Planet-Disk Perturbations in HD~61005 System}
\shortauthors{Esposito et al.}

\begin{document}


\title{Bringing ``The Moth'' to Light: \\A Planet-Sculpting Scenario for the HD~61005 Debris Disk}


\author{Thomas M. Esposito\altaffilmark{1,2}, Michael P. Fitzgerald\altaffilmark{1}, James R. Graham\altaffilmark{2}, Paul Kalas\altaffilmark{2,7}, Eve J. Lee\altaffilmark{2}, Eugene Chiang\altaffilmark{2}, Gaspard Duch\^{e}ne\altaffilmark{2,6}, Jason Wang\altaffilmark{2}, Maxwell A. Millar-Blanchaer\altaffilmark{3,4}, Eric Nielsen\altaffilmark{5,7}, S. Mark Ammons\altaffilmark{8}, Sebastian Bruzzone\altaffilmark{9}, Robert J. De Rosa\altaffilmark{2}, Zachary H. Draper\altaffilmark{11,12}, Bruce Macintosh\altaffilmark{5}, Franck Marchis\altaffilmark{7}, Stanimir A. Metchev\altaffilmark{9,10}, Marshall Perrin\altaffilmark{14}, Laurent Pueyo\altaffilmark{14}, Abhijith Rajan\altaffilmark{13}, Fredrik T. Rantakyr\"o\altaffilmark{15}, David Vega\altaffilmark{7}, Schuyler Wolff\altaffilmark{16}}


\altaffiltext{1}{Department of Physics and Astronomy, 430 Portola Plaza, University of California, Los Angeles, CA 90095-1547, USA}
\altaffiltext{2}{Department of Astronomy, University of California, Berkeley, CA, 94720, USA}
\altaffiltext{3}{Department of Astronomy \& Astrophysics, University of Toronto, Toronto, ON M5S 3H4, Canada}
\altaffiltext{4}{Dunlap Institute for Astronomy \& Astrophysics, University of Toronto, Toronto, ON M5S 3H4, Canada}
\altaffiltext{5}{Kavli Institute for Particle Astrophysics and Cosmology, Stanford University, Stanford, CA 94305, USA}
\altaffiltext{6}{Universit\'e Grenoble Alpes / CNRS, Institut de Plan\'etologie et d'Astrophysique de Grenoble, F-38000 Grenoble, France}
\altaffiltext{7}{SETI Institute, Carl Sagan Center, 189 Bernardo Avenue, Mountain View, CA 94043, USA}
\altaffiltext{8}{Lawrence Livermore National Laboratory, 7000 East Ave., Livermore, CA 94550, USA}
\altaffiltext{9}{Department of Physics and Astronomy, Centre for Planetary Science and Exploration, the University of Western Ontario, London, ON N6A 3K7, Canada}
\altaffiltext{10}{Department of Physics and Astronomy, Stony Brook University, Stony Brook, NY 11794-3800, USA}
\altaffiltext{11}{University of Victoria, 3800 Finnerty Rd, Victoria, BC V8P 5C2, Canada}
\altaffiltext{12}{National Research Council of Canada Herzberg, 5071 West Saanich Rd, Victoria, BC V9E 2E7, Canada}
\altaffiltext{13}{School of Earth and Space Exploration, Arizona State University, PO Box 871404, Tempe, AZ 85287, USA}
\altaffiltext{14}{Space Telescope Science Institute, Baltimore, MD 21218, USA}
\altaffiltext{15}{Gemini Observatory, Casilla 603, La Serena, Chile}
\altaffiltext{16}{Department of Physics and Astronomy, Johns Hopkins University, Baltimore, MD 21218, USA}

\email{tesposito@berkeley.edu}



\begin{abstract}

The HD 61005 debris disk (``The Moth'') stands out from the growing collection of spatially resolved circumstellar disks by virtue of its unusual swept-back morphology, brightness asymmetries, and dust ring offset. Despite several suggestions for the physical mechanisms creating these features, no definitive answer has been found. In this work, we demonstrate the plausibility of a scenario in which the disk material is shaped dynamically by an eccentric, inclined planet. We present new Keck NIRC2 scattered-light angular differential imaging of the disk at 1.2--2.3 $\micron$ that further constrains its outer morphology (projected separations of 27--135 AU). We also present complementary Gemini Planet Imager 1.6 $\micron$ total intensity and polarized light detections that probe down to projected separations less than 10 AU. To test our planet-sculpting hypothesis, we employed secular perturbation theory to construct parent body and dust distributions that informed scattered-light models. We found that this method produced models with morphological and photometric features similar to those seen in the data, supporting the premise of a planet-perturbed disk. Briefly, our results indicate a disk parent body population with a semimajor axis of 40--52 AU and an interior planet with an eccentricity of at least 0.2. Many permutations of planet mass and semimajor axis are allowed, ranging from an Earth mass at 35 AU to a Jupiter mass at 5 AU.

\end{abstract}



\section{INTRODUCTION} \label{sect:61005_intro}

The gravitational influences of massive bodies such as planets and brown dwarfs can shape the spatial distributions of planetesimals and grains in circumstellar debris disks. Observationally, this mechanism is best studied when the disk is spatially resolved on scales small enough to distinguish individual features of the disk morphology. Near-infrared imaging of starlight scattered by a disk's micron-sized dust using large ground-based telescopes provides the necessary resolution and sensitivity, thus making it a powerful tool for investigating disk--planet interaction. Theoretical modeling of the system dynamics can then constrain physical parameters of both planets and disks.

The detection and subsequent modeling of debris disk structures also serve to guide future attempts at planet detection, as exemplified by the $\beta$ Pictoris system \citep{smith1984}. In that case, the detection of a dust-depleted inner region and warped disk indicated the presence of a massive companion (e.g., \citealt{lagage1994, mouillet1997}) over a decade before the giant planet $\beta$ Pic b was first detected \citep{lagrange2009a}. This is not an isolated occurrence, as nearly all of the directly imaged exoplanets to date reside in systems hosting substantial dust disks, some with irregular morphologies (e.g., $\beta$ Pic, Fomalhaut, HR 8799, HD 95086, HD 106906; \citealt{kalas2008, lagrange2010, marois2010_8799, rameau2013_95086b, bailey2014}).

HD 61005 is a young (40-100~Myr; \citealt{desidera2011}), nearby ($\sim$35~pc; \citealt{perryman1997}), G8Vk star.  Mid-infrared (IR), far-IR, and submillimeter observations indicated substantial amounts of dust and larger grains \citep{hillenbrand2008, meyer2008, roccatagliata2009, ricarte2013}. \citet{hines2007} and \citet{maness2009} resolved the disk in scattered light with \emph{Hubble Space Telescope (HST)} NICMOS 1.1~$\micron$ and ACS 0.6~$\micron$ observations, respectively.  The disk viewing geometry was found to be near edge-on, but included a sharp bend in both projected midplanes that led \citet{hines2007} to name HD 61005 ``The Moth'' due to its overall wing-like appearance.  These early observations revealed a surface brightness asymmetry between the two sides of the disk (NE twice as bright as SW), and follow-up, Very Large Telescope (VLT)/NaCo, near-IR angular differential imaging \editbf{(ADI; \citealt{marois2006})} discovered an inner cleared region consistent with a ring inclined by $\sim$84$\degr$ and narrow streamers extending outward from the ring ansae \citep{buenzli2010}. The ring size (radius $\sim$ 61 AU) was consistent with the spectral energy distribution (SED) modeled by \citet{hillenbrand2008}, and a 2.75~AU projected stellocentric offset was also discovered. \emph{HST} STIS optical imaging from \citet{schneider2014} showed a more complete view of the low surface brightness ``skirt'' of dust stretching between the streamers south of the star, seen previously with NICMOS and ACS. Most recently, \citet{olofsson2016} presented high signal-to-noise ratio (S/N) VLT/SPHERE IRDIS $H$- and $K$-band images that further confirm the known features while also showing that the ring brightens with decreasing projected separation and the E ansa remains brighter than the W ansa in polarized intensity. That same work reported Atacama Large Millimeter/submillimeter Array (ALMA) 1.3 mm data indicating that the disk's large grains are confined to a ring with a semimajor axis of $\sim$66 AU.

To date, two different models have been proposed to explain the wing-like morphology. \citet{hines2007} hypothesized that a cloud of gas in the interstellar medium (ISM) could be exerting ram pressure on small grains and unbinding them. \citet{debes2009} produced a disk model based on this hypothesis that roughly approximated the swept-back shape. On the other hand, \citet{maness2009} asserted that the observed line-of-sight gas column density is too low to drive ram pressure stripping of disk grains. Instead, they proposed that the swept-back morphology could be caused by secular (i.e., long-period) perturbations of grains due to gravitational forces exerted by low-density (warm), neutral interstellar gas. However, models based on this mechanism were again only able to roughly reproduce the disk's observed features and did not reproduce the NE/SW brightness asymmetry. Furthermore, there is as yet no observational evidence for warm gas clouds in the vicinity of HD 61005, and, if such a cloud is present, the grain--gas interaction timescale could become too long to significantly shape the disk if the mean gas density is too low or the cloud has a filamentary morphology.

In this work, we introduce a third model that shows that the morphology of the HD~61005 debris disk could result from the secular perturbation of grain orbits due to gravitational interaction with an inclined, eccentric companion. Such a companion could have a range of substellar masses; we will refer to it as the ``planet'' for simplicity. Prior theoretical studies of similar scenarios have shown that eccentric planets can induce stellocentric ring offsets and disk brightness asymmetries on long timescales \citep[e.g.][]{wyatt1999, pearce2014, pearce2015a}. 
To evaluate the role of a putative planet in shaping the HD~61005 debris disk, we adopt a mathematical framework based on the secular perturbation theory described in \citet{wyatt1999}. In addition to eccentricity effects, we include the effects of mutual inclination between the planet and disk. This framework simulates the influence of a planet on circumstellar grains and then constructs 2D scattered-light models of the disk. In parallel, \citet{lee2016} have expanded on this concept to explain the menagerie of observed debris disk morphologies from first principles.

Complementing the models, we present new scattered-light imaging of the disk in the form of Keck NIRC2 ADI $J$-, $H$-, and $K_p$-band data, as well as Gemini Planet Imager (GPI; \citealt{macintosh2014}) polarimetric $H$-band data. We compare our data quantitatively with this model and thus constrain parameters for both the disk and perturber using a Markov-Chain Monte Carlo (MCMC) fitting technique. Additionally, we report photometric and morphological measurements of the disk based on our high angular resolution, multiwavelength imaging.

We provide details about our observations and data reduction methods in Section \ref{sect:61005_obs}, and we present our imaging results in Section \ref{sect:img_results}. We then describe our secular perturbation model and present our model results in Section \ref{sect:pert_model}. Afterward, in Section \ref{sect:61005_discussion}, we discuss the implications of our observational and model results in the contexts of the HD 61005 system and beyond. Finally, we summarize our conclusions in Section \ref{sect:conclusions}.

\section{OBSERVATIONS AND DATA REDUCTION} \label{sect:61005_obs}

\subsection{Keck NIRC2} \label{sect:keck_obs}

We observed HD~61005 on three separate nights between 2008 and 2014 using the Keck II adaptive optics (AO) system and a coronagraphic imaging mode of the NIRC2 camera. Three different broadband filters were used: $J$, $H$, and $K_p$. See Table \ref{tab:obs_specs} for filter central wavelengths, exposure times, numbers of exposures, field rotation per data set, and observation dates. The camera was operated in ``narrow'' mode, with a $10\arcsec\times10\arcsec$ field of view (FOV) and a pixel scale of 9.95 mas $\mathrm{pixel^{-1}}$ \citep{yelda2010}. A coronagraph mask of radius 200 mas occulted the star in all science images. Airmass ranged from 1.62 to 1.67 across the three nights, and the AO loops were closed, with HD 61005 serving as its own natural guide star.

\begin{table}[ht]
\begin{center}
\caption{HD 61005 Observations}
\label{tab:obs_specs}

\setlength{\tabcolsep}{5pt}
\begin{tabular}{l c c c c c l}
\toprule
  Inst. & Filt & $\lambda_{\mathrm{c}}$ & $t_{\mathrm{exp}}$ & $N_{\mathrm{exp}}$ & $\Delta$PA & Date  \\
   &  & ($\micron$) & (s) &  & (deg) &   \\
  \midrule
  NIRC2 & $J$ & 1.25 & 20.0 & 65 & 11.1 & 2014 Feb 09 \\
  NIRC2 & $H$ & 1.63 & 60.0 & 66 & 25.0 & 2008 Dec 02 \\
  \multirow{2}{*}{NIRC2} & \multirow{2}{*}{$K_p$} & \multirow{2}{*}{2.12} & 20.0 & 42 & \multirow{2}{*}{19.9} & \multirow{2}{*}{2012 Jan 03} \\
   & &  & 30.0 & 61 &  &  \\
  GPI & $H$ & 1.65 & 59.6 & 35 & 140.7 & 2014 Mar 24 \\
  \bottomrule
\end{tabular}

\end{center}
\end{table}

For calibration purposes, we observed standard stars FS 123, FS 155, and FS 13 \citep{hawarden2001} unocculted to determine the photometric zeropoint in $J$, $H$, and $K_p$ bands, respectively. Conditions were photometric on each night. Flux densities used for flux conversion were taken from \citet{tokunaga2005}.
    
We employed ADI for all science observations. This technique fixes the telescope point-spread function's (PSF) orientation relative to the camera and AO system optics during the observations. As a result, the FOV rotates throughout the image sequence, while the PSF orientation remains constant relative to the detector.

We used the same preliminary reduction procedure for all three data sets. After bias subtraction and flat-fielding, we masked cosmic-ray hits and other bad pixels. Next, we aligned the individual exposures via cross-correlation of their stellar diffraction spikes \citep{marois2006}. Following this, radial profile subtraction and a median boxcar unsharp mask (box width 40 pixels) served as high-pass filters to suppress the stellar halo and sky background.

Following the procedure described in \citet{esposito2014}, we applied a modified LOCI algorithm (``locally optimized combination of images''; \citealt{lafreniere2007}) to suppress the stellar PSF and quasi-static speckle noise in our $H$ and $K_p$ data. For each image in a data set, LOCI constructs a unique reference PSF from an optimized linear combination of other images in the data set. In azimuthally divided subsections of stellocentric annuli, the coefficients $c_{ij}$ of the linear combination are chosen so as to minimize the residuals of the PSF subtraction. To simplify the ADI self-subtraction forward-modeling that we apply to our models (\citealt{esposito2014}, see Sections 2 and 3.4), our algorithm computes the median of the $c_{ij}$ across all subsections containing disk signal in a given annulus (known from preliminary reductions) and replaces the original $c_{ij}$ with that median value. After PSF-subtracting our images, we derotated them, masked residuals from the diffraction spikes, and then mean-collapsed the image stack to create the final images presented in Figure \ref{fig:imstack}.

We tuned the LOCI parameters manually to achieve a balance between noise attenuation and disk flux retention. Following the conventional parameter definitions from \citet{lafreniere2007}, we used values of 
$W$=10 pix, $N_{\delta}$=0.1, $dr$=10 pix, $g$=0.1, and $N_a$=500
for both the $H$ and $K_p$ data.

Although we preferred LOCI because we were better able to characterize the self-subtraction bias it introduces, it performed poorly on the $J$-band data set, so we employed \texttt{pyklip}, a Python implementation \citep{wang2015_pyklip} of the Karhunen--Lo\`eve Image Projection (KLIP) algorithm \citep{soummer2012, pueyo2015}. In this process, we divided the images into stellocentric annuli, divided each annulus into azimuthal subsections, and computed the principal components of each subsection. The main parameters we adjusted were the number of modes used from the Karhunen--Lo\`eve (KL) transform and an angular exclusion criterion for reference PSFs similar to $N_\delta$. The images were then derotated and mean-combined into the final image shown in Figure \ref{fig:imstack}. The KLIP parameters used in this reduction are as follows: 20 annuli between $r=21$ and 400 pixels, no azimuthal division of the annuli (i.e., one subsection per annulus), a minimum rotation threshold of 1$\arcdeg$, and projection onto one KL mode (the primary mode only). We did not mask the diffraction spikes at any point. This reduction also returned a higher-S/N result than a basic ADI reduction in which a median of all images composed the reference PSF.

\editbf{Reductions of the $H$ data with \texttt{pyklip} produced lower S/N than LOCI due to greater attenuation of disk brightness, particularly in the disk's wings. Combined with LOCI's advantages in characterizing self-subtraction bias, this led us to choose LOCI over KLIP in this case (and for NIRC2 $K_p$ for consistency during analysis).}

\subsection{Gemini Planet Imager} \label{sect:gpi_obs}

GPI is a high-contrast imager on the 8 m Gemini South telescope with a high-order, natural guide star AO system \citep{macintosh2014, poyneer2014_gpi} to correct for atmospheric turbulence, a coronagraph that suppresses starlight, and an integral field unit (IFU) for spectroscopy and broadband imaging polarimetry \citep{larkin2014_gpi}. The AO correction allows near diffraction-limited imaging over a $\sim2.7\arcsec\times2.7\arcsec$ FOV. GPI always observes in an ADI mode. HD~61005 was observed during instrument verification and commissioning in 2014 March. Table \ref{tab:obs_specs} details the observations. The instrument was operated in its $H$-band polarimetry mode, with a pixel scale of 14.166 $\pm$ 0.007 mas $\mathrm{lenslet^{-1}}$ \citep{derosa2015b}. A 123 mas radius coronagraph mask occulted the star in all science images. Airmass ranged from 1.008 to 1.003 during the observations.

The Wollaston prism used in polarimetry mode splits the light from the IFU's lenslets into two orthogonal polarization states, producing two spots per lenslet on the detector. To reduce these data, we used the GPI Data Reduction Pipeline (\citealt{perrin2014_drp}) and largely followed the reduction methods described in \citet{perrin2015_4796} and \citet{millar-blanchaer2015}, which we summarize here. The raw data were dark subtracted, flexure-corrected using a cross-correlation routine, fixed for bad pixels in the 2D data, and assembled into data cubes containing both polarization states using a model of the polarimetry mode lenslet PSFs. These cubes were then corrected for distortion \citep{konopacky2014}, corrected for noncommon path biases between the two polarization spots via double differencing, and fixed for bad pixels in the 3D data cube. At this point, we smoothed the images using an FWHM $=2$ pixel Gaussian profile, subtracted the estimated instrumental polarization, and aligned them using measurements of the four fiducial diffraction or ``satellite'' spots, which are centered on the location of the occulted star \citep{wang2014, pueyo2015}. The resulting data cubes were rotated to place north along the $+y$-axis and they were all then combined using singular value decomposition matrix inversion to obtain a three-dimensional Stokes cube containing the Stokes parameters \{$I$, $Q$, $U$, $V$\}. Finally, the data were photometrically calibrated using the satellite spot fluxes and an HD 61005 flux of $H=6.578$ mag (Two Micron All Sky Survey [2MASS]) as described in \citet{hung2015b}.

\begin{figure*}[ht!]
\centering
\includegraphics[width=6.0in]{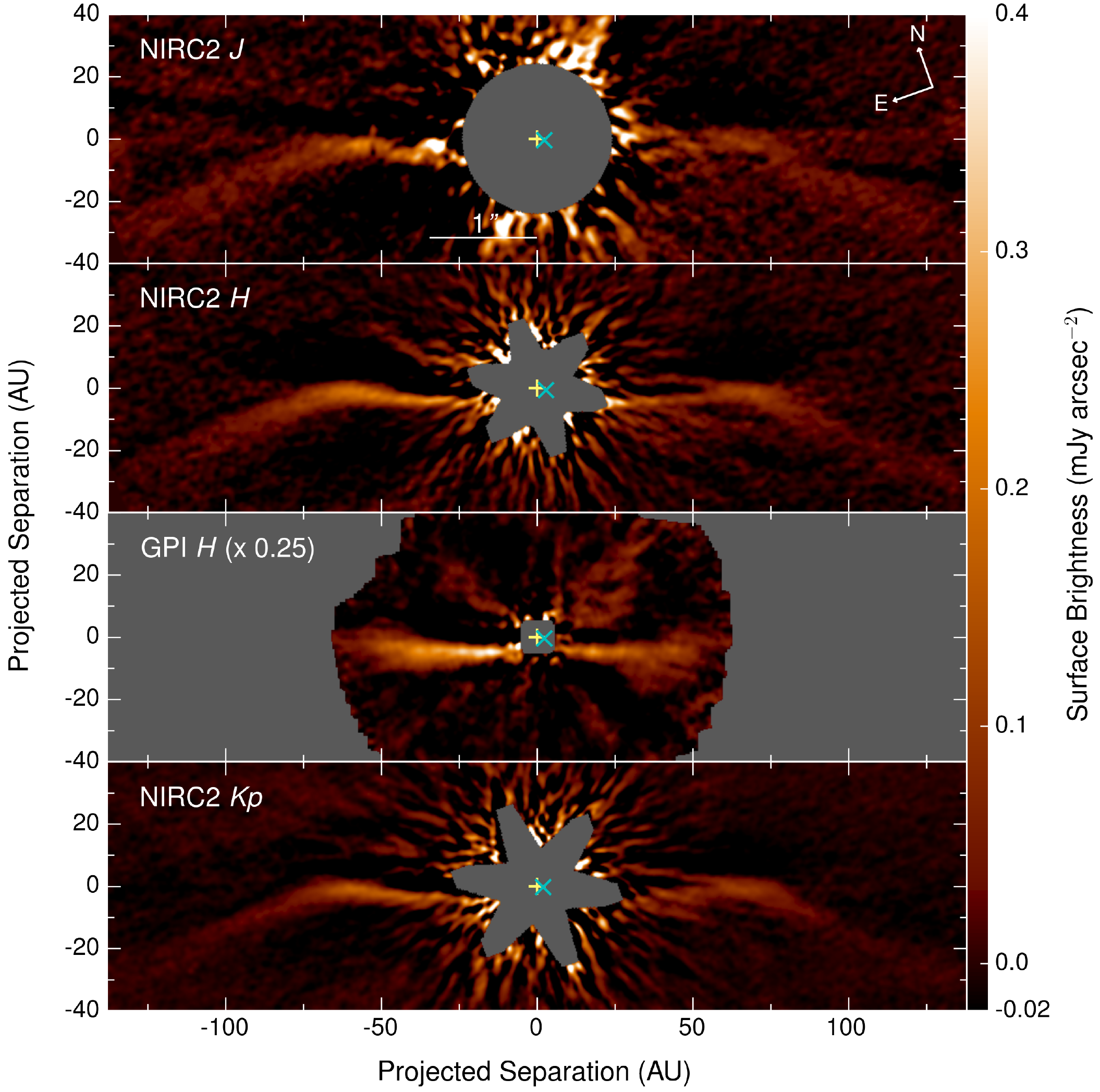}
\caption{PSF-subtracted images of the disk's scattered-light surface brightness, rotated with the major axis horizontal. For display only, images were smoothed after final combination with a $\sigma=2$ pixel Gaussian kernel. Panels 1, 2, and 4 (from top) show Keck NIRC2 data in $J$, $H$, and $K_p$ bands. The $H$-band data contain the highest S/N, while the $J$-band data suffered from limited field rotation and coincidence of the disk PA with the Keck diffraction spikes. The swept-back wings and the east-to-west and front-to-back brightness asymmetries are clear in all three bands, while the inner clearing and ring center (cyan cross) offset from the star (yellow plus sign) are seen in $H$ and $K_p$. The ring center could not be accurately measured in the NIRC2 $J$ and GPI images, so the center position marked there is simply a mean of the NIRC2 $H$ and $K_p$ centers. Panel 3 shows KLIP-reduced GPI $H$-band total intensity data, scaled in brightness by a factor of 0.25 for display purposes. The GPI data show the brightness asymmetries and inner clearing, but the field of view did not encompass the wings.}
\label{fig:imstack}
\end{figure*}

To subtract the stellar PSF from the total intensity (Stokes $I$) images, we used the same \texttt{pyklip} algorithm as for the NIRC2 $J$-band data. The final image shown in Figure \ref{fig:imstack} was created using 30 annuli evenly spaced between $r=6$ and 135 pixels, five azimuthal subsections per annulus, a minimum rotation criterion of 10$\arcdeg$ for allowed reference images, and 11 KL modes.

\section{HIGH-CONTRAST IMAGING RESULTS} \label{sect:img_results}

We present PSF-subtracted scattered-light images of the disk from NIRC2 in the $J$, $H$, and $K_p$ bands and a GPI $H$-band total intensity image in Figure \ref{fig:imstack}. The images were rotated $19\fdg3$ counterclockwise so that the disk's major axis lies horizontal, and they were smoothed (after PSF subtraction and combination) by a Gaussian kernel with standard deviation $\sigma=2$ pixels for display only. We spatially resolve the disk at projected separations of $\sim$27--135 AU ($0\farcs79$--$3\farcs88$) with NIRC2 and $\sim$9--51 AU ($0\farcs26$--$1\farcs48$) with GPI. Interior to these regions, disk emission is obscured by the focal plane mask and contaminated by residual speckle noise. Exterior to these regions, the disk signal approaches the background level for NIRC2 and is truncated by the limited FOV of GPI. Negative-brightness regions appear above and below the disk as a result of self-subtraction by LOCI and KLIP processing. The limited field rotation and coincidence of the disk position angle (PA) with the Keck diffraction spikes in the $J$-band data resulted in substantial PSF residuals and a reduced S/N compared to the other two NIRC2 images. Therefore, we report the $J$-band detection but do not include it in our detailed analyses. We also detect the disk in polarized intensity with GPI, which we discuss more in Section \ref{sect:pol}.

\subsection{Disk Morphology} \label{sect:61005_morphology}

We detect all of the major morphological features reported previously for this disk: the swept-back wings, stellocentric offset, and inner clearing. The measured PA of the disk's projected major axis is $70\fdg7\pm0.8$ east of north. We measured this in the NIRC2 $H$ and $K_p$ images as the angle of a line connecting the apparent inflection points of the ring's inner edge (i.e., intersection between front and back edges) on both sides of the star. The uncertainty is dominated by a measurement error of $\sim0\fdg8$ ($\pm2$ pixels) in our assumed position of the inflection point (the instruments' systematic errors are $\lesssim0\fdg1$). Both images agree on this value, which is consistent with PAs from previous publications, and the $J$ and GPI images (with no clear inflection point) are visually consistent as well.

\textit{Wings}: the swept-back wings are detected with NIRC2 but lie outside of GPI's FOV. They show a sharp bend at the ring ansae like an ``elbow,'' with deflection angles of $\sim$22$\arcdeg$ on the E side and $\sim$25$\arcdeg$ on the W (measured relative to the ring's major axis by manually tracing the brightest pixels in the wing at each separation). This $\sim$3$\arcdeg$ difference is consistent between the $H$ and $K_p$ images, suggesting that it is a real feature. Measuring outward from the elbows in $H$ band, the wings extend from $\sim$62 to 127 AU ($1\farcs79$--$3\farcs70$) on the E side and from $\sim$67 to 135 AU ($1\farcs94$--$3\farcs88$) on the W. Their extents are similar in $J$ and $K_p$. The stellocentric offset is evidenced by the $\sim$5~AU difference in inner extent for the two wings. The difference in outer extent is more difficult to interpret, as the disk's surface brightness reaches our sensitivity limit and we likely do not see the true endpoints of the wings.

\textit{Ring Offset}: we measured the center of the ring to be offset from the star in NIRC2 $H$ by $2.5\pm0.8$~AU to the W along the major axis and by $0.6\pm0.5$~AU to the S along the minor axis. Similarly, we measured an offset in NIRC2 $K_p$ of $1.9\pm0.8$~AU to the W and $0.3\pm0.5$~AU to the S. To measure the ring center, we fit ellipses to the NIRC2 $H$ and $K_p$ rings after aggressively high-pass filtering the images to leave only the highest spatial frequency components of the ring. The uncertainties are the quadrature sum of Gaussian $1\sigma$ uncertainties from the least-squares fit ($\sim$1 and 2 pixels in minor and major, respectively) and the estimated uncertainty in the absolute star position behind the focal plane mask ($\pm$1 pixel in $x$ and $y$). The spatially extended ansae lead to larger uncertainties along the major axis than the minor axis. The $H$ and $K_p$ measurements are statistically consistent with each other and with the $2.75\pm0.85$~AU offset to the W reported by \citet{buenzli2010}. Residuals from the diffraction spikes and speckles in the $J$ image interfered with ellipse fitting, as did the limited FOV of the GPI image. Therefore, we do not report offsets for those data and plot the ring center for those images in Figure \ref{fig:imstack} as the mean of the NIRC2 $H$ and $K_p$ centers merely for reference.

\textit{Inner Clearing}: the disk's dust appears to be depleted inside of the ring in our images. This is consistent with findings by \citet{buenzli2010} and \citet{schneider2014}. We note that ADI self-subtraction may artificially suppress disk brightness inside of the ring \citep{milli2012}. However, our detection of both the front and back edges of the ring is evidence of a true deficit in brightness, and thus dust, rather than just a reduction artifact. As we discuss later, our modeling also supports this interpretation (see Section \ref{sect:model_results}).

\subsection{Disk Photometry} \label{sect:photometry}

In all of our images, the ring's south edge is substantially brighter than the north edge. Based on an assumption of primarily forward-scattering grains constituting an optically thin disk, we consider the brighter edge to be the front edge (i.e., closer to the observer). The W side of the back edge is weakly detected in NIRC2 $H$ and $K_p$ and is undetected in the other images. We do not detect the E back edge at all, even in conservative reductions. On the other hand, the ring's E front edge is $\sim$1.5--2.5 times brighter than the W front edge at similar projected separations, which is consistent with previous resolved imaging of the disk. This pattern holds even at the smallest separations seen with GPI.

\begin{figure}[ht]
\centering
\includegraphics[width=\columnwidth]{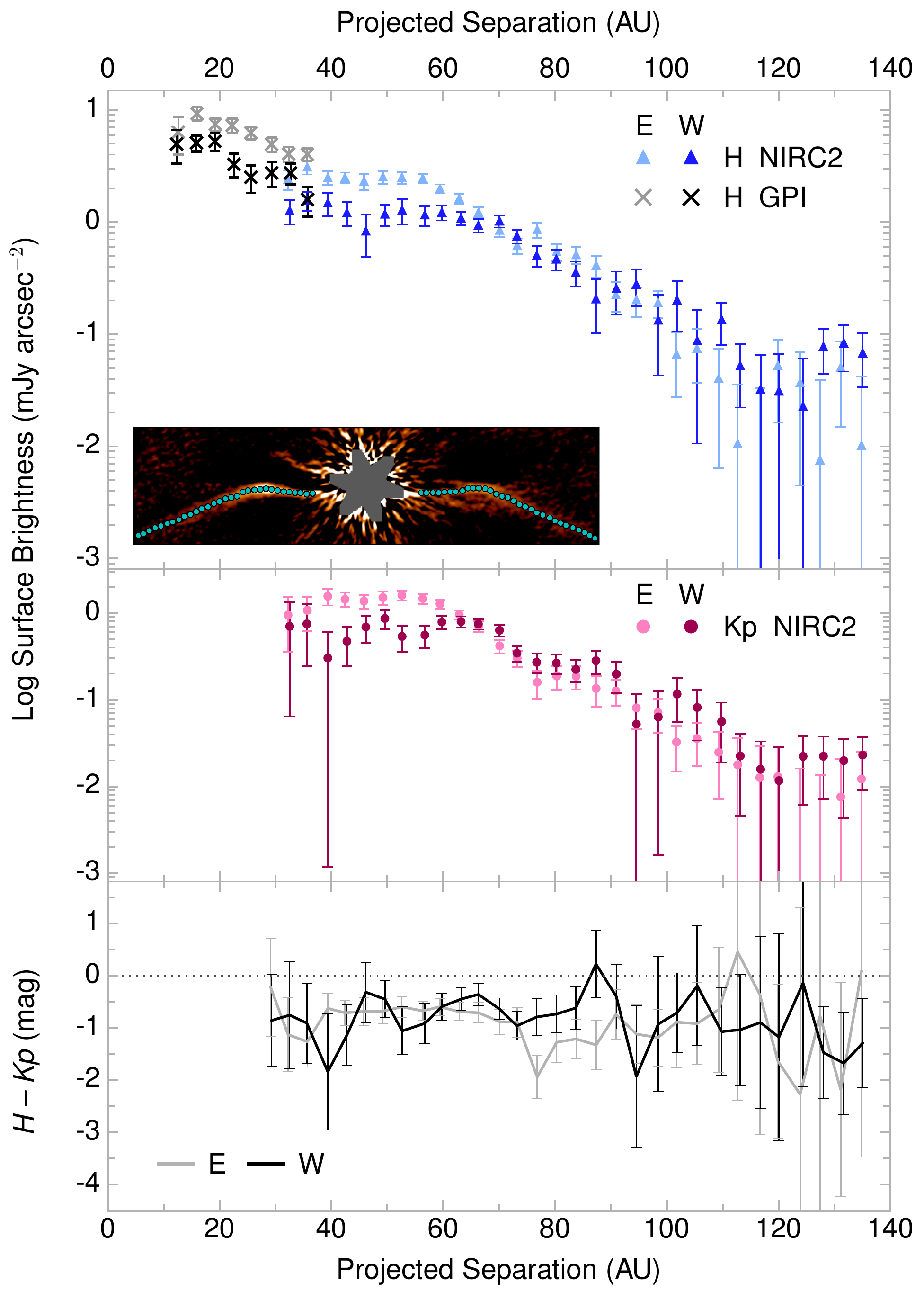}
\caption{\textit{Top:} surface brightness radial profiles for both sides of the disk (E and W) in $H$ band with NIRC2 (blue) and GPI (gray). We measured the mean surface brightnesses inside circular apertures of radius 5 pixels placed along the ring's front edge and the wings (see inset) and applied an ADI self-subtraction correction. The ring is brighter to the E than the W, but the wings are symmetric. \editbf{The ring ansae appear as shoulders at $\sim$55 AU (E) and $\sim$65 AU (W) beyond which there are breaks in the profile slope.} GPI measurements show the ring growing continuously brighter as separation decreases. \textit{Middle:} NIRC2 $K_p$ profiles with the same general features as $H$ but with systematically lower brightnesses. \textit{Bottom:} the disk's $H-K_p$ color after subtracting the star's color. The disk is consistently blue in all regions, suggesting scattering dominated by submicron-sized grains.}
\label{fig:sb_color}
\end{figure}

To quantify some of these brightness features, we measured surface brightness radial profiles for the disk by performing aperture photometry on $H$ (NIRC2 and GPI) and $K_p$ (NIRC2) reduced images. The results are plotted in the top two panels of Figure \ref{fig:sb_color}. The NIRC2 profiles were measured from the images in Figure \ref{fig:imstack}, while the GPI profiles were measured from a KLIP-reduced total intensity image (Figure \ref{fig:polstack}) that was designed to conserve more disk brightness than the reduction shown in Figure \ref{fig:imstack}, using 20 annuli, three subsections, minimum rotation of 8$\arcdeg$, and three KL modes. Circular apertures 5 pixels in radius were placed along the ring's front edge and wings at discrete projected radii and centered on the peak of the emission in that region (see Figure \ref{fig:sb_color} inset). These apertures are smaller than the width of the ring at its narrowest point, and thus we expect them to only include disk brightness and not artificial negative brightness created by ADI self-subtraction.

The raw profiles are still biased by self-subtraction in the processed images, however, so we divide each aperture's surface brightness by a correction factor. For the NIRC2 data, we first computed a ratio of the raw and self-subtracted disk models presented in Section \ref{sect:model_results}. The self-subtracted models were forward-modeled using the NIRC2 $H$ and $K_p$ LOCI parameters following the procedure described in \citet{esposito2014}. Correction factors were then estimated at each aperture location as the mean measured inside the aperture in the ratio image. For the GPI correction, we injected a fake disk into the individual frames at a PA rotated 90$\arcdeg$ relative to the real disk and re-reduced those data using KLIP with the same parameters as the original reduction. We then computed the correction factors as the ratios of the unprocessed fake disk's brightnesses to the KLIP-processed fake's brightnesses, similar to the NIRC2 procedure. The NIRC2 correction factors ranged from 1.2 to 5.2 and the GPI factors ranged from 1.5 to 2.3, with the larger factors at smaller separations.

To estimate the uncertainties on these measurements, we first calculated the mean brightnesses within many ``pure-noise'' apertures located at the same separation as the disk measurement but well outside the disk. We then took the standard deviation of those means and added it in quadrature with the estimated photon noise for the measured disk brightness. Finally, we scaled this sum by the self-subtraction correction factor for the aperture in question.

In $H$ and $K_p$, the ring's brightness is greatest close to the star and decreases with separation out to the ansae, although the innermost measurements have large uncertainties due to stellar PSF residuals and extreme self-subtraction bias. The GPI data in particular highlight this trend, which is also clear in SPHERE data \citep{olofsson2016}. Farther out, the ansae appear as flat shoulders in the radial profiles \editbf{beyond which there is a break in the profile slope}. In each filter, the ring and ansae are brighter in the E than the W by a factor of $\sim$2. The \editbf{break} associated with the W ansa is also shifted farther from the star than the E ansa. This is possibly due to the ring offset, which can account for a shift of $\sim$6 AU (twice the measured stellocentric offset). The offset of the \editbf{breaks} is almost 10 AU, however, so other factors may be affecting the disk brightness. We expand on this in Section \ref{sect:61005_disc_obs}.

In contrast to the ring, there is no significant brightness asymmetry in the wings. The wing brightness also generally decreases with separation but does so at a slower rate than seen in the ring.

\subsection{Disk Color} \label{sect:61005_color}

We calculated the disk's $H-K_p$ color based on the NIRC2 surface brightness radial profiles shown in the top panel of Figure \ref{fig:sb_color} and present it in the bottom panel of that figure. The host star's color was calculated from 2MASS measurements \citep{cutri2003} and subtracted from the disk color. The mean color of the disk, weighted by the measurement uncertainties, over all separations (29--135 AU) is $-0.96$ mag and $-0.94$ mag E and W of the star, respectively. This makes the disk distinctly blue.

To check whether different regions of the disk displayed different colors, we calculated the weighted means for three ranges in projected separation: interior to the ansae ($<$52 AU), within the ansae (52--76 AU), and exterior to the ansae (in the wings, $>$76 AU). These means, in magnitudes, for the (E,W) sides of the disk are interior = (-0.89, -1.02), ansa = (-1.00, -0.62), exterior = (-1.21, -0.93). Therefore, the blue color is approximately constant with projected separation and consistent between the two sides of the disk.

\subsection{Disk Polarization} \label{sect:pol}

\begin{figure}[ht]
\centering
\includegraphics[width=\columnwidth]{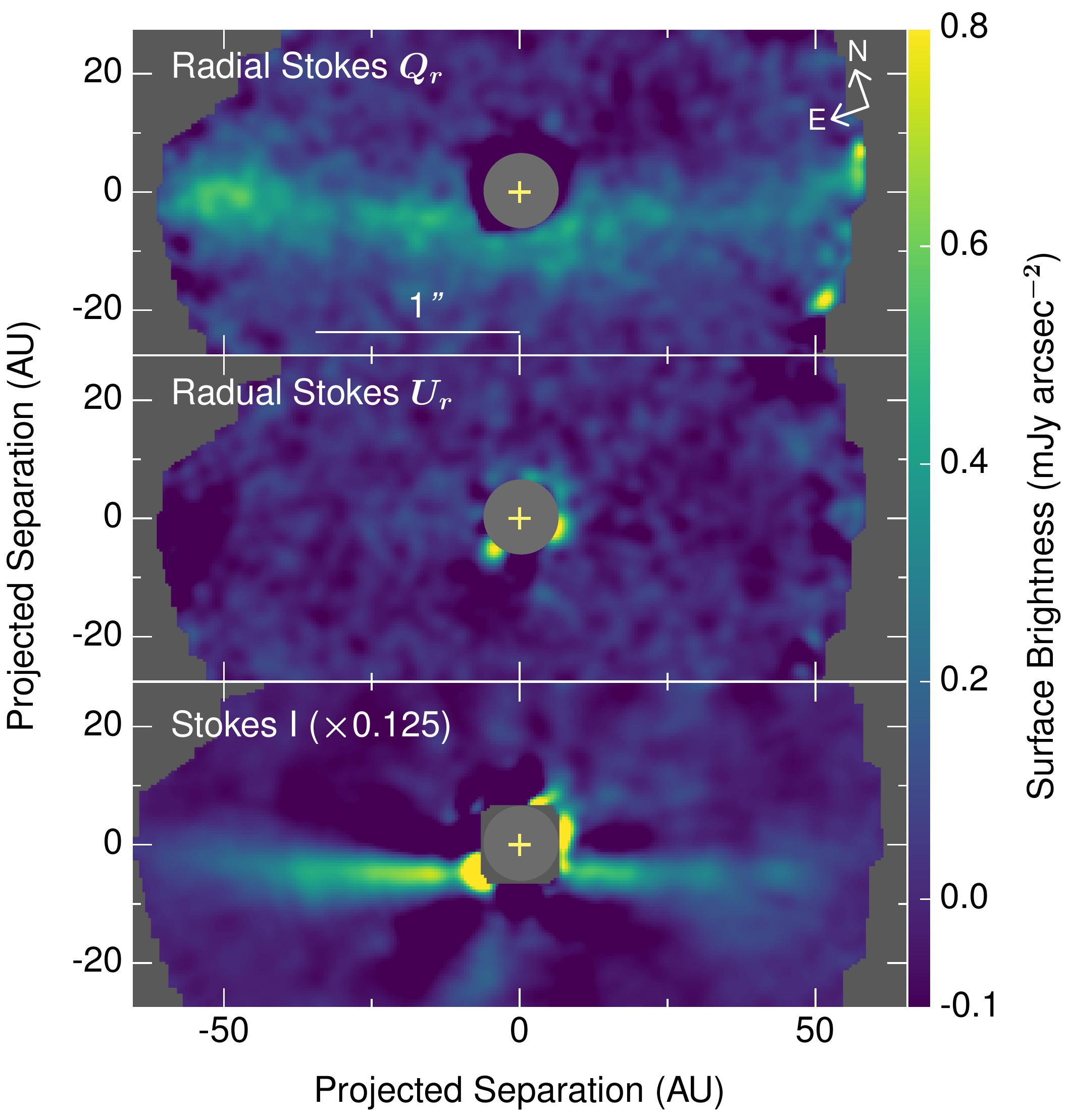}
\caption{\textit{Top:} the disk in the radial Stokes $Q_r$ polarization state with GPI. \textit{Middle:} the disk in radial Stokes $U_r$, effectively representing a noise map because we do not expect significant scattering by circumstellar material in this polarization state. \textit{Bottom:} GPI Stokes \textit{I} (total intensity) via a conservative KLIP reduction, scaled down by a factor of 8 for display. All images are in $H$ band, were rotated so that the disk's projected major axis is horizontal, and were smoothed (after all reduction steps) with a $\sigma=2$ pixel Gaussian kernel for display only. The star is marked by a yellow plus sign.}
\label{fig:polstack}
\end{figure}

We detected the disk in linearly polarized light with GPI, shown in the top two panels of Figure \ref{fig:polstack}. To facilitate analysis, we transformed GPI's Stokes $Q$ and $U$ polarization components into their more intuitive radial analogs, $Q_r$ and $U_r$ \citep{schmid2006}. $Q_r>0$ indicates a polarization vector perpendicular to a line drawn from the star to the pixel in question, while $Q_r<0$ indicates a polarization vector parallel to such a line. $U_r$ is analogous to $Q_r$, but the polarization vectors are rotated by $\pm$45$\arcdeg$. We do not expect single scattering by circumstellar material to generate a significant $U_r$ signal, so we treat the $U_r$ image as a noise map for $Q_r$.

The brightness asymmetries seen in total intensity persist in the disk's $Q_r$ brightness, with the E side still $\sim$2 times brighter than the W and the front of the ring brighter than the back. There is also no discernible signal from the back side of the ring. On the other hand, the disk appears more extended vertically in polarized light than in total intensity at the same separation, with the polarized disk almost twice as wide at some separations. This is likely an effect of ADI processing filtering out some of the low-frequency signal in total intensity, with no such effect in polarized intensity from PDI.

\begin{figure}[h]
\centering
\includegraphics[width=\columnwidth]{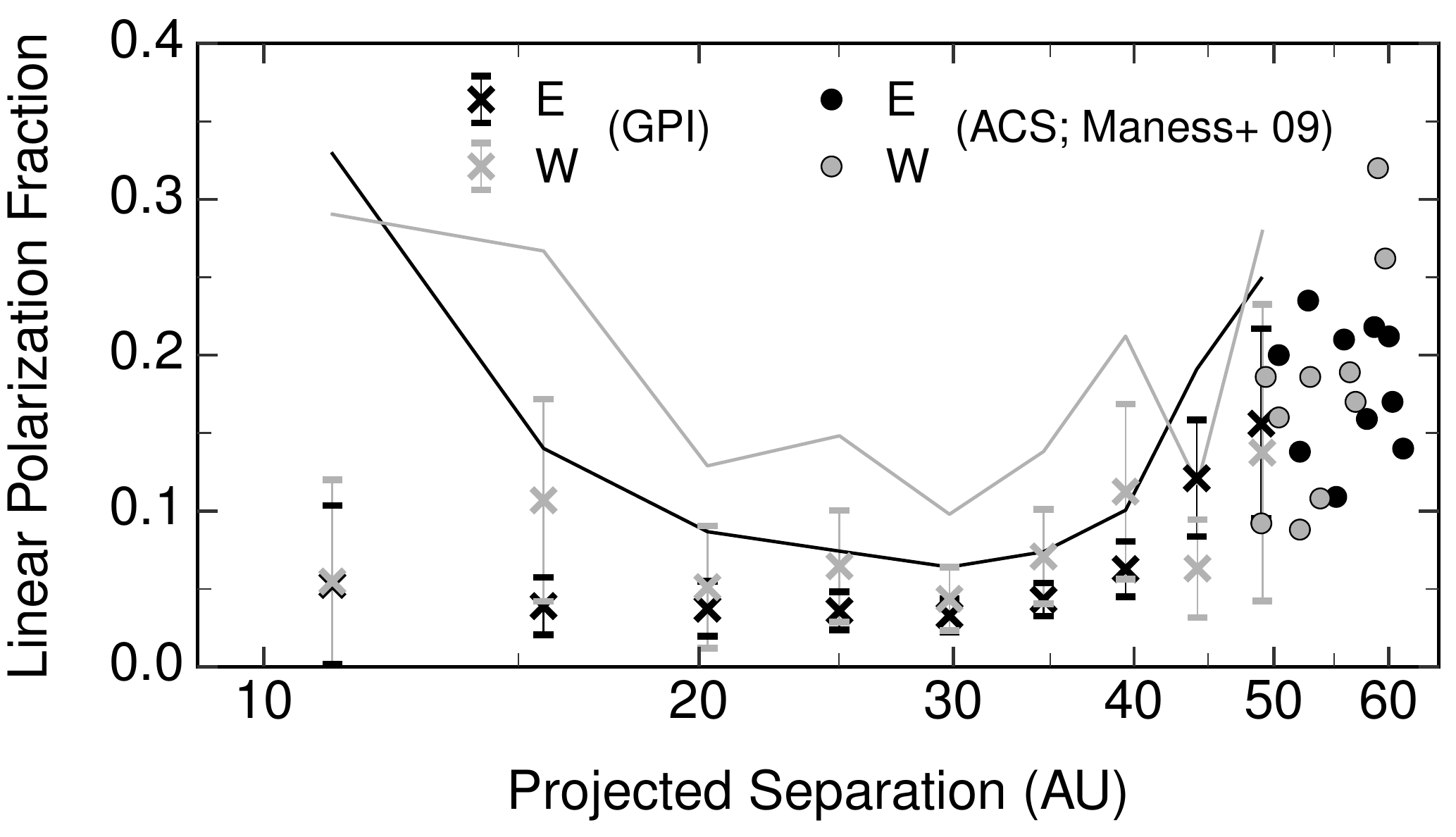}
\caption{Fractional linear polarization of the disk measured with GPI. The linear polarization fraction is calculated as $Q_r/I$ (the radial Stokes $Q$ divided by self-subtraction-corrected total intensity). For comparison, we include polarization fraction measurements from $HST$ ACS 0.6 $\mu$m data presented by \citet{maness2009} (adapted from their Figure 4). The polarization fraction is only a few percent interior to 35 AU but shows an increasing trend exterior to that point. Error bars on the GPI points represent pure measurement uncertainty, while the solid lines represent upper limits based on uncertainty in the self-subtraction correction.}
\label{fig:pol_frac}
\end{figure}

We also computed the total linear polarization fraction as a function of projected separation (Figure \ref{fig:pol_frac}). We measured the mean fraction within circular apertures 5 pixels in radius centered on the disk's brightest pixel at a given separation. This fraction is calculated as $Q_r$/$I$, with $I$ being the total intensity. We exclude $U_r$ from the polarized intensity because it would introduce additional noise and bias the quantity. To mitigate the effects of self-subtraction bias on the polarization fraction, we measured $I$ from the conservative KLIP reduction of the GPI total intensity described earlier (bottom panel of Figure \ref{fig:polstack}) and corrected it for self-subtraction in the same manner as the surface brightness profiles.

The total linear polarization fraction is consistently $\sim$4\%--7\% at 11--35 AU, then increases to $\sim$15\% at $r\approx48$ AU. This fraction is similar for the two sides of the disk. Our values are in rough agreement with those that \citet{maness2009} found for the polarization fraction at $\lambda$ = 0.6 $\mu$m. Their measurements did not extend inward of 48 AU, but they are in line with our values in this region. Additionally, they found the polarization fraction to follow a positive power law as a function of distance from the star (index$\approx$0.1) that, if extrapolated, would approach our measurements of $\sim$5\% at the innermost separations. 

\subsection{Sensitivity to Companions} \label{sect:companions}

\begin{figure}[h]
\includegraphics[width=\columnwidth]{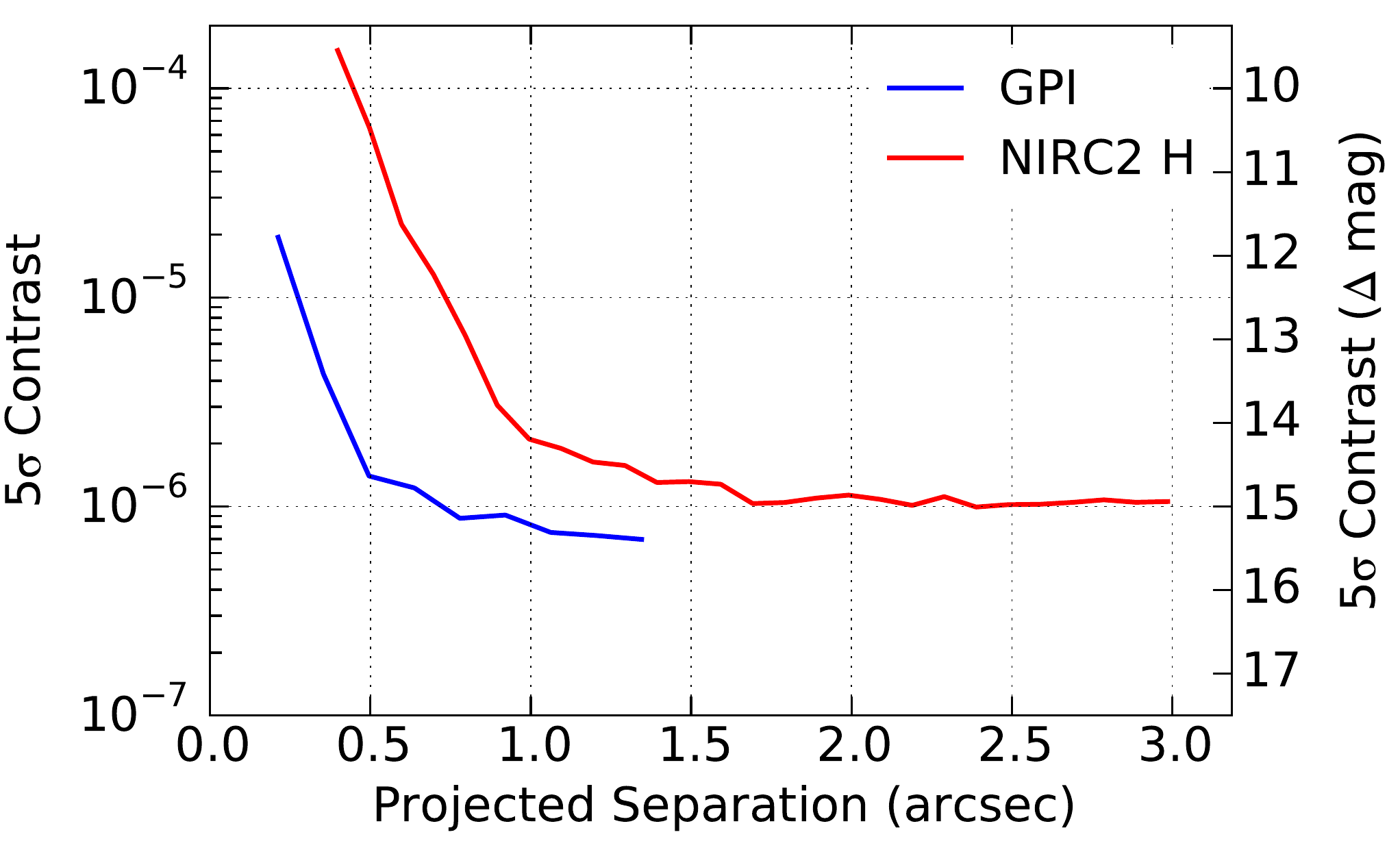}
\includegraphics[width=\columnwidth]{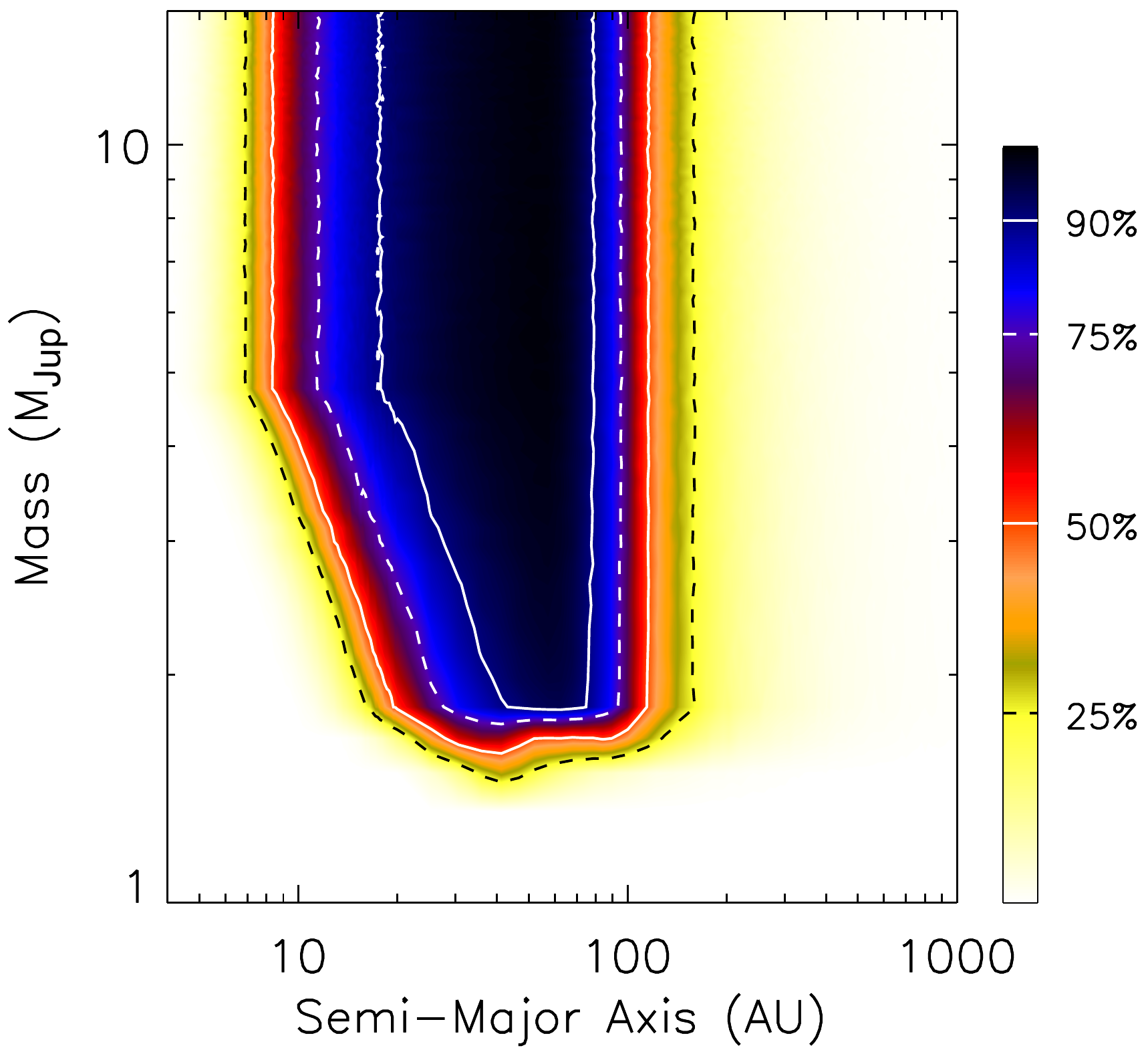}
\caption{\textit{Top:} the 5$\sigma$ equivalent false positive thresholds for point-source companions (i.e., contrast curves) from our GPI total intensity and NIRC2 $H$-band images. \textit{Bottom:} our observational completeness for companions as a function of mass and semimajor axis based on those 5$\sigma$ thresholds.}
\label{fig:contrast}
\end{figure}

\editbf{Although we did not detect any companions to HD 61005, we can constrain potential companion masses and semimajor axes based on the sensitivity of our observations. We determined the 5$\sigma$ equivalent false positive thresholds for point sources, or contrast curves, for our GPI total intensity and NIRC2 $H$-band images following the method outlined in \citet{mawet2014} and plotted them in Figure \ref{fig:contrast}. It is important to note that we mask the disk when measuring contrast; thus, our contrasts and the resultant completeness estimates are overestimated for regions in the disk. Flux attenuation due to PSF subtraction was quantified and corrected by injecting and recovering the brightnesses of simulated planets.}

\editbf{Following the procedure used by \citet{wang2015_aumic}, we translated our contrast curves to limits on possible companions by running a Monte Carlo analysis as described by \citet{nielsen2008} and \citet{nielsen2010} to determine the completeness of our data. In this process, planets with random orbits are generated (including randomized inclination), the contrast curves for both epochs determine whether the planets are detected, and COND atmosphere models \citep{baraffe2003} are used to convert from planet luminosity to mass. The limits are reported in Figure \ref{fig:contrast}, with the colors and contours denoting completeness to planets of given mass and semimajor axis, assuming that their flux is not conflated with disk flux.}

\section{MODELING DISK SECULAR PERTURBATIONS} \label{sect:pert_model}

We investigate a scenario in which an unseen planet on an inclined,
eccentric orbit perturbed the disk's grains secularly. In the
following sections, we construct models of the disk in scattered
light and compare them to a subset of the data presented above.

\subsection{Model Overview} \label{sect:moddesc}

Here we describe how we construct 
models of secularly perturbed disks 
and their images in scattered light.
The secular perturbation theory behind our model is described in
detail by \citet{wyatt1999}. We summarize the components of that
theory relevant to our model and refer the reader to the original
publication for further details.

Particles that constitute our model debris disks 
consist of two types: ``parent bodies'' and ``dust grains.''
The latter spawn through collisional fragmentation of the former.
A planet, embedded in the disk, secularly perturbs 
the parent bodies (the planet's gravitational potential 
is treated as a massive wire; see, e.g., \citealt{murray_dermott}).
Each particle is characterized by its orbital semi-major axis $a$,
eccentricity $e$, inclination $I$, longitude of ascending node
$\Omega$, longitude of pericenter $\widetilde{\omega}$, and $\beta$,
the ratio of the stellar radiation pressure to stellar gravity.

A parent body's $e$, $I$, $\widetilde{\omega}$, and $\Omega$ can be
broken down into proper and forced elements.
The proper elements (denoted by subscript p) are the particle's
``intrinsic'' elements, i.e., those that the particle would have if
there were no perturber in the system. The forced elements (denoted
by subscript f) are contributed by the perturber and depend 
on its orbital elements, as well as 
the ratio between the perturber's and the particle's semi-major
axes. With only one perturber in the system, the forced elements
imposed on a particle are constant in time and
independent of the perturber's mass. 
Our calculations account only for linear secular perturbations and
not those of mean motion resonances. Secular perturbations do not
change semi-major axes.
We ignore disk gravity and assume the perturbing planet's orbital
elements (denoted by subscript ``per'') to be constant on timescales
longer than the precession and collision timescales of the parent
bodies.

The parent bodies are large ($\beta\ll 1$) compared to dust grains and are assumed to have a much smaller
collective surface area; their contribution
to scattered light is neglected.
Throughout this manuscript, variables lacking subscripts belong to
these parent bodies. 
Parent bodies experience collisions and are fragmented into smaller
dust particles (denoted by subscript d). \editbf{In reality, we expect fragmentation to occur at all positions along an orbit over time. To simplify our model, however, we assume that the parents fragment only at periastron, where particle mean velocities are largest and violent collisions may occur more frequently (for an exploration of other assumptions about the orbital
phase of fragmentation, see \citealt{lee2016}).}
Dust particles are assumed to inherit the same orbital velocities
as their parents at the time of breakup; at the same time, the dust
will have a larger $\beta$ due to its smaller size. The result is
that the dust has $I_{d}=I$, $\Omega_{d} = \Omega$, and
$\widetilde{\omega}_{d} = \widetilde{\omega}$, while $a_{d}$ and $e_{d}$
differ from the parent's values according to $\beta$ (see Equations
\ref{eq:a_d} and \ref{eq:e_d}; note that these expressions assume that the dust particles are born at the parent body periastron). In
another simplification, we do not secularly perturb the orbits of
the dust particles after they are born. This is justified because
the collision and blowout lifetimes for the smallest dust are much
shorter than secular perturbation timescales. We also ignore the
effects of Poynting--Robertson drag (see \citealt{wyatt2005a} and
\citealt{strubbe2006}).

\subsection{Model Parameterization}

\begin{figure*}[ht]
\centering
\includegraphics[width=6.5in]{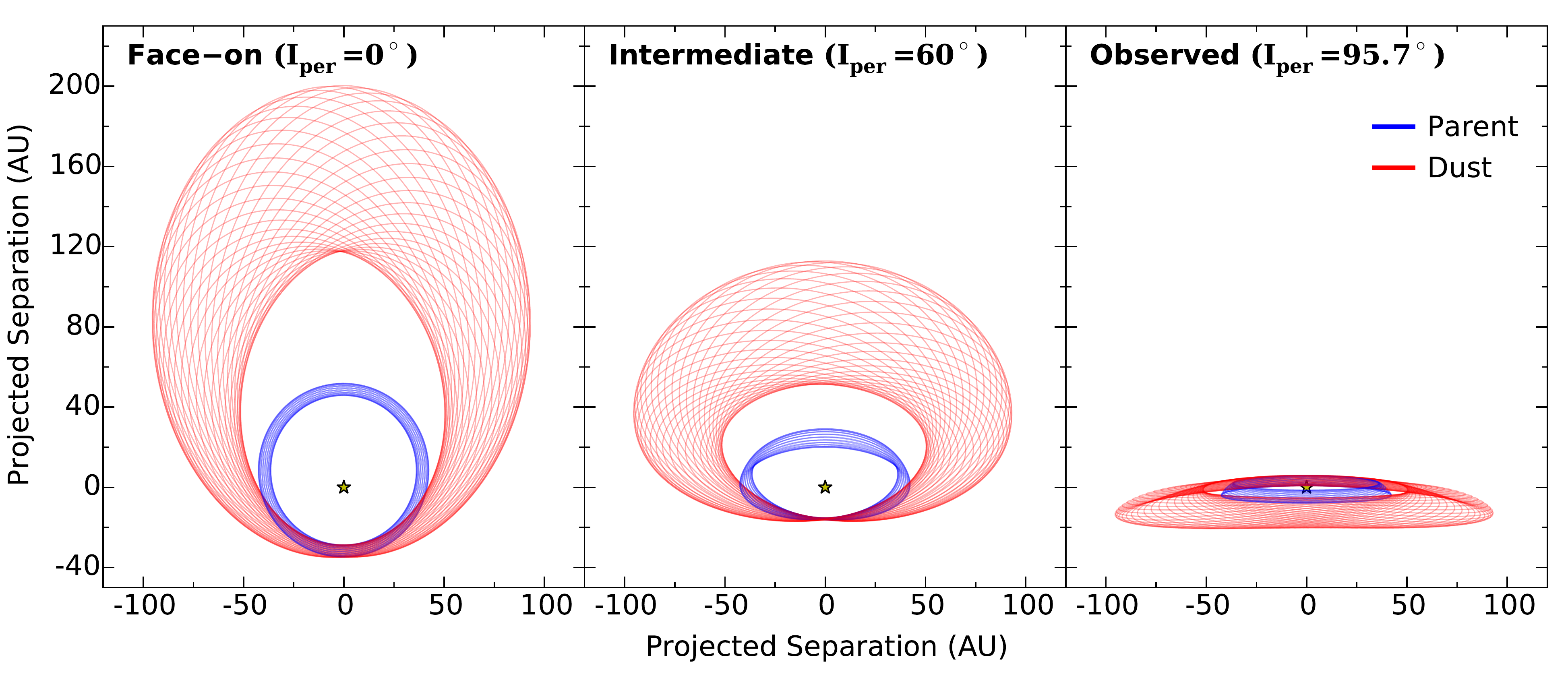}
\caption{The ensemble of parent body (blue) and dust orbits (red) traced out
by our best-fit model. At periastron, the parent bodies break up into the
smaller dust particles that compose our final 3D dust distribution and from
which we derive our scattered-light models (see Figure \ref{fig:modstack}).
The three panels show the same orbits viewed at apparent inclinations of $I_{per}=0\arcdeg$, $I_{per}=60\arcdeg$, and the best-fit model inclination of $I_{per}=95\fdg7$. The star is located at the origin. Only 20 parent body orbits and 40 dust grain orbits are displayed here.}
\label{fig:grain_orbits}
\end{figure*}

We choose the sky plane as the reference plane for the planet's orbit, with the origin coincident with the star. The reference frame is defined such that when the planet's orbit is viewed face on, the on-sky azimuthal coordinate $\theta$ 
is measured counterclockwise from
the downward direction (see Figure \ref{fig:grain_orbits}).

We search for a set of planet and disk parameters that provides the
best-fit model to HD~61005. The planet's orbital inclination
$I_{per}$, argument of periastron $\omega_{per}$, and longitude of
ascending node $\Omega_{per}$ are free parameters. The planet's
eccentricity and semi-major axis are encapsulated by the forced
eccentricity $e_{f}$ of parent bodies, and so we use this last
quantity as a free parameter and not the former two.

``Initial'' values for parameters that have yet to undergo precession
are denoted by subscript ``0.'' The disk's parent bodies are all
assigned the same semimajor axis $a$, initial proper eccentricity 
$e_{p0}$, and initial orientation angles $I_{p0}$, $\omega_{p0}$, and
$\Omega_{p0}$ measured relative to the planet's orbital plane.
Although for simplicity we formally adopt a single semimajor axis $a$ for
all our (80) parent bodies, we account implicitly for a range
of semimajor axes by allowing the parent bodies
to have different ``final'' nodal and periastron longitudes,
i.e., we allow the parent bodies to differentially precess
according to their semimajor axes, which differ in reality.
The precise distribution of nodal and periastron longitudes
will be fitted to the data, as described below when we introduce
our cubic spline function.

The initial (pre-precession) parent body total eccentricity and inclination are given by their complex values $z_0$ and $y_0$, respectively. Each is composed of forced and proper elements:
\begin{equation}
z_0 = z_f + z_{p0}
\end{equation}
\begin{equation}
y_0 = y_f + y_{p0}.
\end{equation}

The forced $z_f$ and $y_f$ are constant:

\begin{equation}
z_f= e_f \mathrm{e}^{i\widetilde{\omega}_{per}}  \label{eq:zf} \\
\end{equation}
\begin{equation}
y_f= I_f \mathrm{e}^{i\Omega_{per}}=0.  \label{eq:yf} \\
\end{equation}
\noindent where $\widetilde{\omega}_{per}=\omega_{per} + \Omega_{per}$. 
The complex forced inclination is zero because we define the parent body inclination relative to the planet's orbital plane; consequently, $I_f=0$. The complex initial proper eccentricities and inclinations for the parent bodies can be written in a similar manner:
\begin{equation}
z_{p0}= e_{p0}\mathrm{e}^{i\widetilde{\omega}_{p0}}  \label{eq:zp} \\
\end{equation}
\begin{equation}
y_{p0}= I_{0}\mathrm{e}^{i\Omega_{0}}  \label{eq:yp} \\
\end{equation}

Now we precess differentially the proper components from the initial values. The proper
eccentricity and inclination for the parent bodies are precessed by $\phi$, an angle that runs
uniformly from 0 to $2\pi$ from our first to our last parent body:
\begin{equation}
z_p=z_{p0}\mathrm{e}^{i\phi}  \\
\end{equation}
\begin{equation}
y_p=y_{p0}\mathrm{e}^{-i\phi}.  \\
\end{equation}
Crucially, the parent bodies do not all carry the same weight when we
compute their contribution to the scattered-light images.  The weight
of a given parent body precessed by $\phi$ is given by a cubic spline
function, $CS (\phi)$, parameterized by six coefficients $b_{[1-6]}$
whose values (together with those of 11 other model parameters) are
adjusted to best fit the images. The cubic spline function is periodic
in $\phi$.  Variations in $CS$ reflect the degree to which parent
bodies have differentially precessed, which in turn depends on how
much they differ in semimajor axis, the mass and \editbf{semimajor axis} of the
perturber, and the age of the system.  Our model is not necessarily
secularly relaxed, i.e., it is not necessarily in steady state.  No
differential precession between parent bodies would make the $CS$ a
delta function. Full differential precession (complete phase mixing)
would make $CS$ constant with $\phi$.

\begin{figure*}[t]
\centering
\includegraphics[width=6.5in]{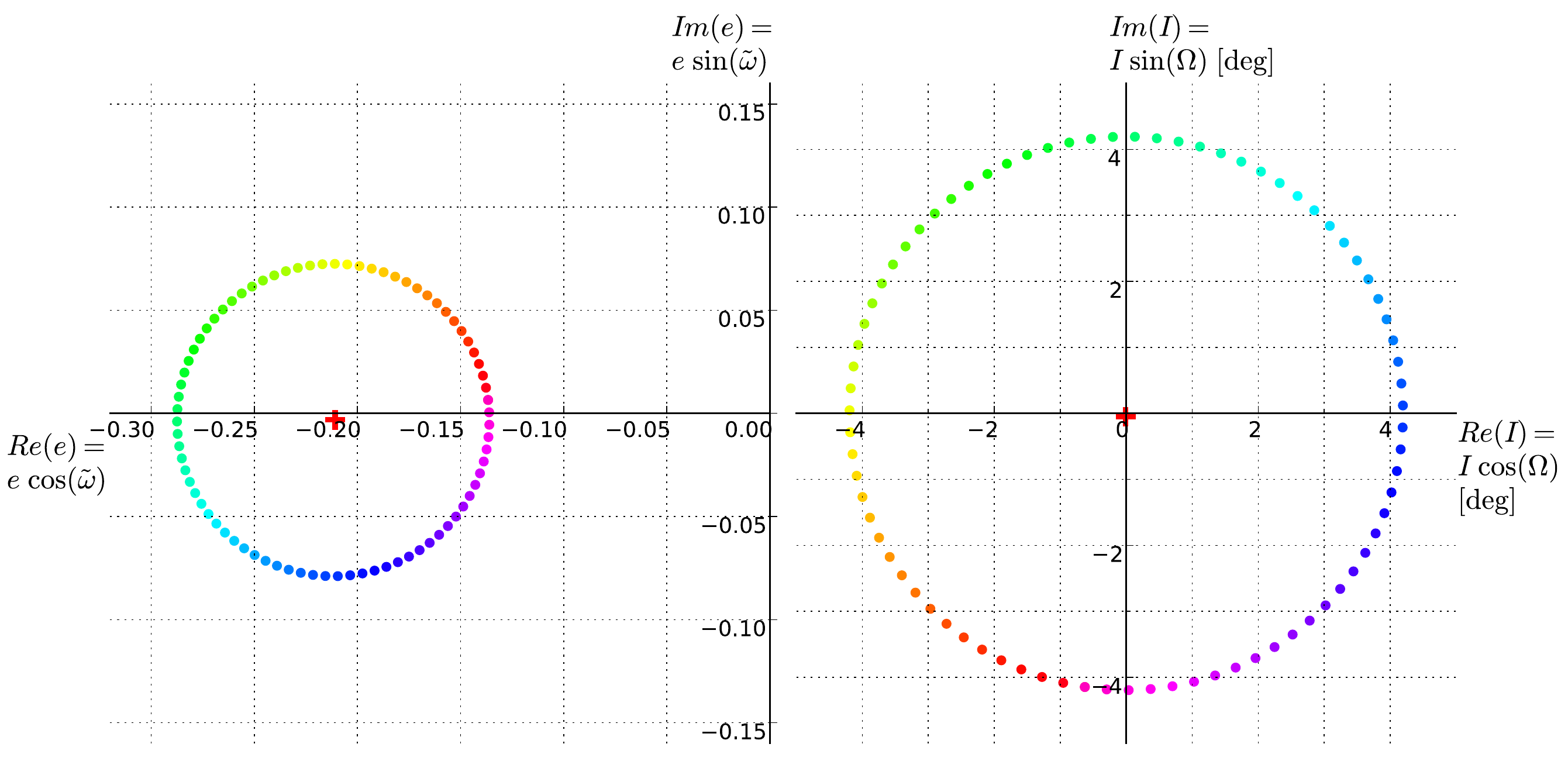}
\caption{Complex eccentricities $z$ (\textit{left}) and inclinations $y$ (\textit{right}) of all 80 parent body orbits, after precession, in the best-fit model. Each parent body orbit is assigned a color that is consistent between panels. The red crosses mark the mean (forced) values.}
\label{fig:parent_e_i}
\end{figure*}

\begin{figure}[h]
\centering
\includegraphics[width=\columnwidth]{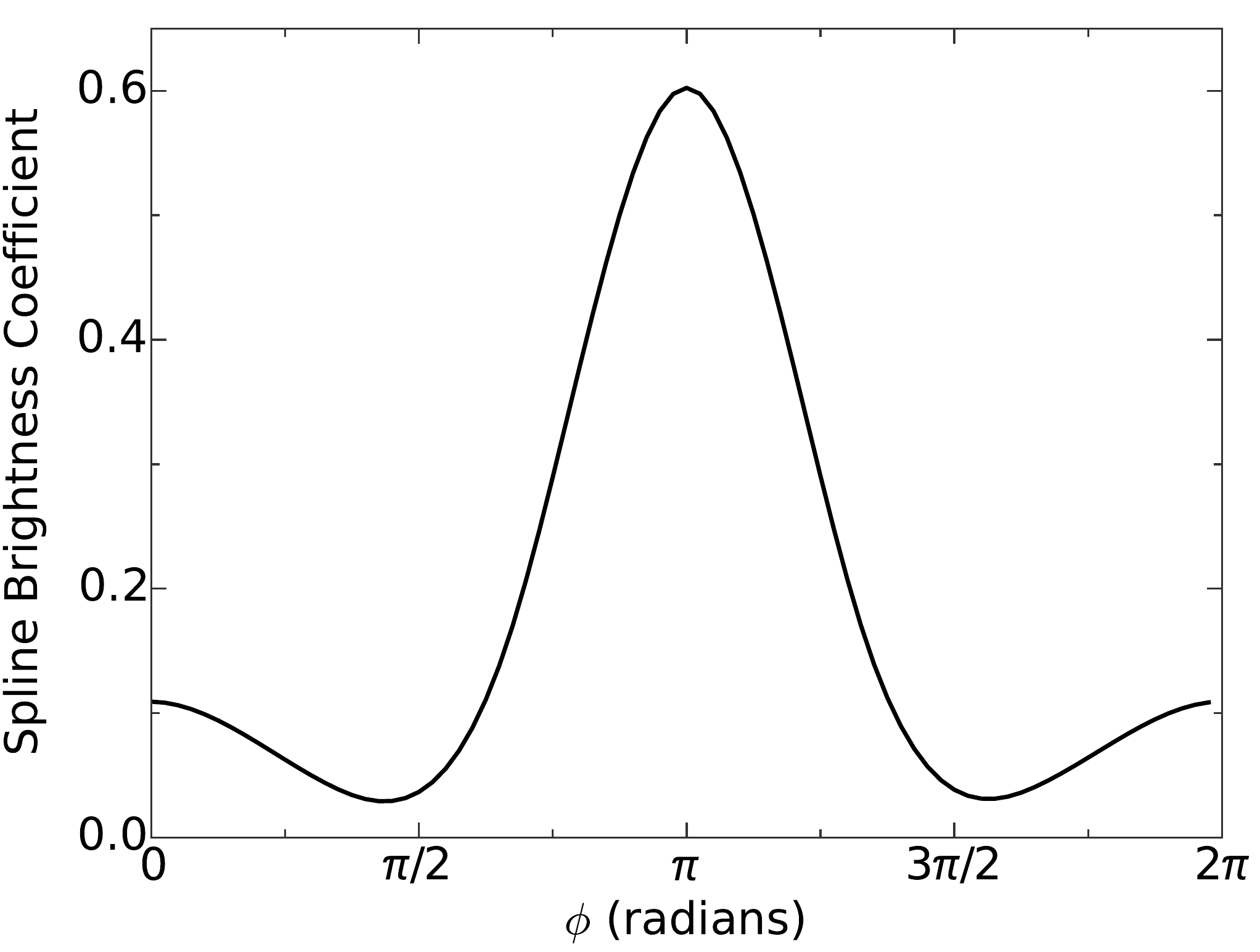}
\caption{The cubic spline function $CS$ that is used to weight the
relative contributions of parent
bodies as a function of the precession angle $\phi$. For the best-fit
model (shown here), the asymmetric weighting is such
that there are more dust particles with low scattering angles on the E
side of the disk (thereby increasing the number of forward-scattered
photons on that side) and fewer particles on the W side. This enhances the
E--W brightness asymmetry of the best-fit model.}
\label{fig:spline}
\end{figure}

Using the precessed proper and forced complex components,
we calculate the total final eccentricity and inclination of each
parent body:
\begin{equation}
z = z_f + z_p  \label{eq:z} \\
\end{equation}
\begin{equation}
y = y_f + y_p  \label{eq:y} \\
\end{equation}
\noindent and take their moduli to obtain the parent body total eccentricity and inclination:
\begin{equation}
e=|z|  \label{eq:e_g} \\
\end{equation}
\begin{equation}
I=|y|.  \label{eq:I_g} \\
\end{equation}
The precessed orbital angles of the parent body are
\begin{equation}
    \Omega = \tan^{-1}[Im(y)/Re(y)]
\end{equation}
\begin{equation}
    \omega = \tan^{-1}[Im(z)/Re(z)] - \Omega.
    \label{eq:omega_EJL}
\end{equation}
We generate 80 parent body orbits following Equation (\ref{eq:e_g}) through
(\ref{eq:omega_EJL}). See Figure \ref{fig:parent_e_i}
for an example set of their total complex eccentricities and inclinations,
and Figure \ref{fig:spline} for an example $CS(\phi)$
(both taken from our best-fit model).

From each parent body is created a dust grain orbit, computed by assuming that dust grains are born exclusively at parent bodies' periastra:
\begin{equation}
a_{d}=\frac{a(1-\beta)}{1-2\beta(1+ e)/(1- e^2)}  \label{eq:a_d} \\
\end{equation}
\begin{equation}
e_{d}=\frac{e + \beta}{1-\beta}.  \label{eq:e_d} \\
\end{equation}
As discussed earlier, $I_{d}$=$I$, $\Omega_{d}$=$\Omega$, and $\widetilde{\omega}_{d}$=$\widetilde{\omega}$. Examples of dust orbits are drawn in red in Figure \ref{fig:grain_orbits}, along with parent body orbits in blue. 

We distribute 200 dust grains evenly spaced in mean anomaly $M$ along each orbit so that
$80\times200 = 16,000$ dust particles contribute to our scattered light images. Dust grains
scatter light according to a Henyey--Greenstein phase function:
\begin{equation}
B_{HG} = \frac{1-g^2}{(1+g^2-2g\cos\theta_{sc})^{3/2}} \label{eq:HG} \\
\end{equation}
where $g$ is the asymmetry parameter (a
free parameter we fit) and $\theta_{sc}$ the scattering angle.
The contribution of each dust
grain scales as $B_{HG}(\theta_{sc}) \times CS(\phi) / r^2$, where $r$
is the distance between the grain and the star.

The result is a ``raw'' model of the disk's scattered-light surface brightness as projected onto the sky and free of ADI self-subtraction. At this point, we smooth the model with a Gaussian kernel with $\sigma=1$ binned pixel (4 NIRC2 pixels) to mitigate artifacts from finite particle number and to approximate diffraction-limited seeing at 1.6 $\mu$m. 

\subsection{MCMC Model Fitting} \label{sect:model_fitting}

We used a parallel-tempered MCMC simulation to explore the 17-variable parameter space that our
model encompassed and find a best-fit model to our data. This simulation was run using the Python module \texttt{emcee} \citep{foreman-mackey2013} with 20 temperatures, 250 walkers, 4150 steps
per walker, and a 350-step ``burn-in'' period on 64 cores of UCLA's Hoffman2 Cluster. 

\begin{figure*}[ht]
\centering
\includegraphics[width=6.0in]{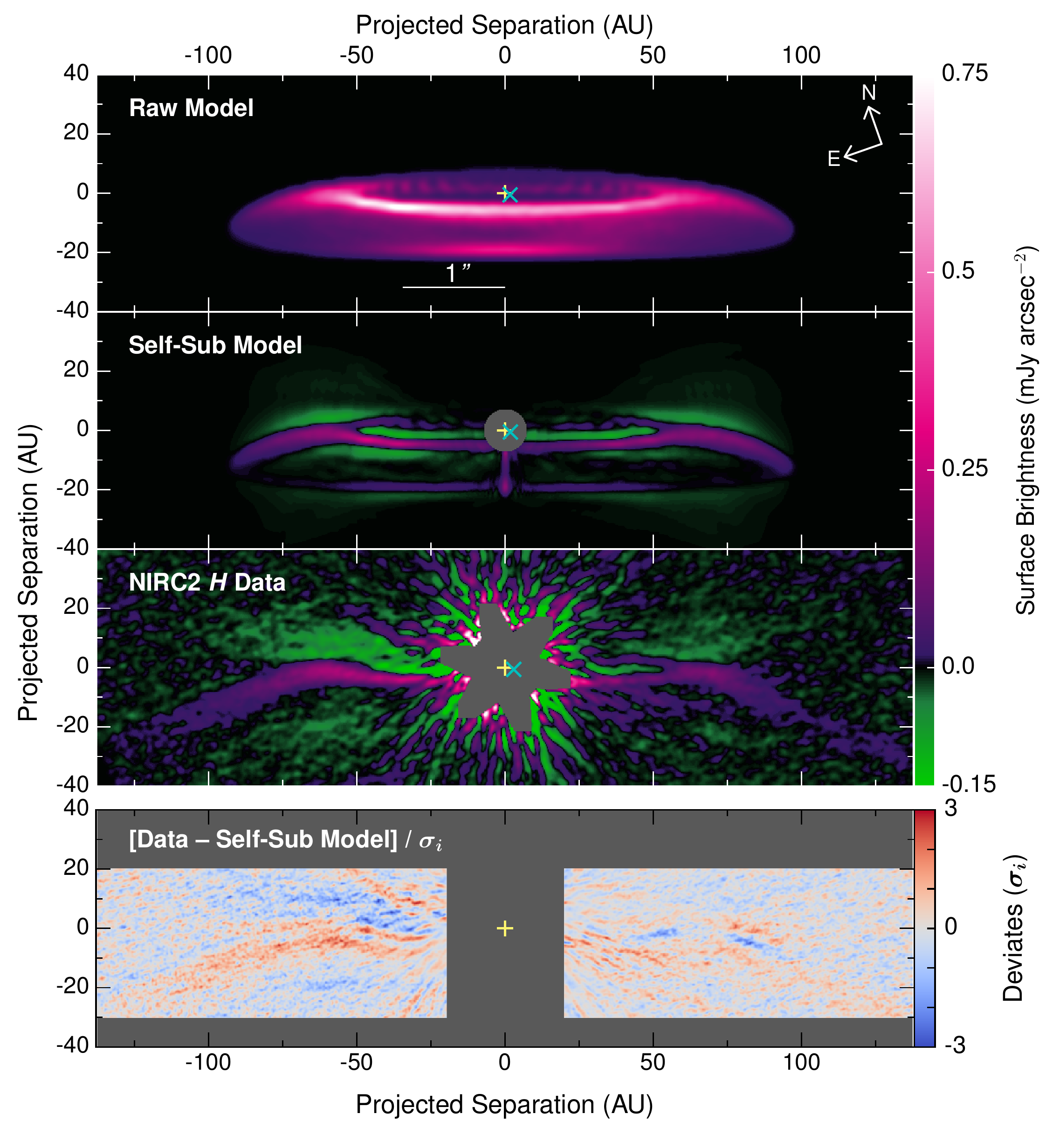}
\caption{Our best-fit model of the scattered-light surface brightness from the eccentric, inclined perturber scenario, compared with the data. Panels from top to bottom: the raw model representing the disk as it would appear before processing-induced biases; the same model after LOCI self-subtraction forward-modeling was applied; the LOCI-processed NIRC2 $H$-band data; and a deviate map. We calculate the deviate map as (data $-$ self-subtracted model)/$\sigma$ where $\sigma$ is the estimated surface brightness uncertainty at each pixel. The swept-back wings, E $>$ W and front $>$ back brightness asymmetries, inner clearing, and ring center (cyan cross) offset from the star (yellow plus sign) are reproduced by our model. These features are particularly emphasized by LOCI self-subtraction. The raw model has not been filtered by ADI image-processing and shows an apron of dust south of the star, which is consistent with previous space-based observations and the GPI data.}
\label{fig:modstack}
\end{figure*}

We chose to fit models only to the $H$-band NIRC2 data because we were primarily interested in modeling the extended wing structures that GPI did not capture. Additionally, we found similar morphologies and brightness relationships in all three NIRC2 images, so to save computation time, we elected to fit only to the image with the highest S/N, which was the $H$-band image. We also binned the data into 4 $\times$ 4 pixel bins (approximately the size of one resolution element). This had the advantages of reducing spatial correlation between adjacent pixels and reducing computation time.

The model that was actually fit to the data was a self-subtracted version of the raw model produced using the method described above. Using the LOCI parameters from the $H$-band reduction and our forward-modeling algorithm from \citet{esposito2014}, we applied self-subtraction to the raw surface brightness model. This resulted in a model biased analogously to the data.

Both the model and data comprised many pixels that contained only noise. Including these pixels in the fits wasted computation time and biased $\chi^2$ downward, so we masked out these regions and excluded them from the calculation of the residuals. The masked pixels are gray in the deviate map of Figure \ref{fig:modstack} (bottom panel). After masking, the weighted residuals for each fit were calculated at each pixel as \textit{res}=(\textit{``data'' $-$ ``self-subtracted model''})/$\sigma$, where $\sigma$ is the brightness uncertainty at that pixel. We calculated $\sigma$ for each pixel $p$ as the standard deviation of the mean brightnesses within apertures at the same separation from the star as $p$ but avoiding disk signal or self-subtracted negative brightness. Therefore, $\sigma$ is the same for each pixel at a given separation.

To find the model that best agreed with the data, we initially performed a coarse grid search
over a wide range of possible parameter values. From there, we manually tuned parameters until we
arrived at a model that roughly resembled the data. To further refine the fit,
we first tried a Levenberg--Marquardt least-squares algorithm but found that it became mired in
local $\chi^2$ minima. Ultimately we performed an MCMC simulation, the results of which
are discussed below.

\vspace{5mm}
\subsection{Modeling Results} \label{sect:model_results}

The MCMC simulation returned parameter values that produce a model similar to the observed disk in many respects. The best-fit (i.e., maximum likelihood) model from the simulation is shown in the top panel of Figure \ref{fig:modstack}, and the second panel from the top shows that model with self-subtraction forward-modeling applied. It is this self-subtracted model that was compared with the data (reproduced in third panel from top) with a reduced chi-squared of $\chi_\nu^2=1.14$. The bottom panel shows the best-fit model's deviate map, calculated at each pixel as (\textit{``data'' $-$ ``self-subtracted model''})/$\sigma$. Table \ref{tab:bf_params} lists the parameter values associated with the best-fit model (i.e., maximum likelihood, $L_{max}$) and the values for the samples in the marginalized distributions corresponding to the 16th, 50th, and 84th percentiles. The 50th percentile value represents the median, and the 16th and 84th percentile values are akin to $\pm 1$ $\sigma$ uncertainties (were the marginalized distributions Gaussian).

\begin{table}[ht]
\begin{center}
\caption{MCMC Model Parameters}
\label{tab:bf_params}
\begin{tabular}{l c c c c c} 
\toprule
Param. & $L_{max}$ & 16\% & 50\% & 84\% & Unit  \\ 
\midrule


$a$ & 40.4 & 42.5 & 44.6 & 52.1 & AU \\ 
$I_{per}$ & 95.7 & 95.3 & 95.6 & 95.9 & deg \\ 
$I_0$ & 4.2 & 3.8 & 4.1 & 4.2 & deg \\ 
$e_f$ & 0.21 & 0.23 & 0.25 & 0.27 & ... \\ 
$e_{p0}$ & 0.08 & 0.08 & 0.09 & 0.14 & ... \\ 
$g$ & 0.60 & 0.58 & 0.59 & 0.60 & ... \\ 
$\beta$ & 0.26 & 0.18 & 0.23 & 0.25 & ... \\
$\Omega_{per}$ & 277.5 & 272.7 & 277.3 & 280.9 & deg \\ 
$\Omega_{p0}$ & 261.4 & 261.6 & 264.2 & 265.9 & deg \\ 
$\omega_{per}$ & 263.3 & 258.5 & 261.6 & 267.0 & deg \\ 
$\omega_{p0}$ & 106.0 & 101.1 & 104.3 & 108.3 & deg \\ 
$b_1$ & 0.026 & 0.015 & 0.034 & 0.052 & ... \\ 
$b_2$ & 0.151 & 0.094 & 0.139 & 0.176 & ... \\
$b_3$ & 0.023 & 0.004 & 0.018 & 0.036 & ... \\
$b_4$ & 0.007 & 0.006 & 0.021 & 0.046 & ... \\
$b_5$ & 0.899 & 0.787 & 0.847 & 0.912 & ... \\
$b_6$ & 0.009 & 0.010 & 0.035 & 0.075 & ... \\

\bottomrule
 
\end{tabular}
\end{center}
\end{table}


At first glance, the model contains all of the features of the observed disk when we qualitatively compare the self-subtracted model to the data. It has swept-back wings that extend outward from the ring ansae at angles. There is a narrow dust ring with an inner clearing and center offset from the star. The brightness of the ring's front edge is greater than that of the back edge, and the E side of the disk is brighter than the W.

The raw model contains those same features but also shows significant
emission south of the star in an apron similar to that seen in the non-ADI STIS data of \citet{schneider2014}. This is important because it means that our model produces a three-dimensional dust distribution consistent with both ground-based and space-based observations, despite only being fit to the former. One STIS-detected feature that we do not reproduce in our model is the pair of spurs extending radially outward from the ring ansae. Conversely, our raw model has a loop of dust north of the ring's back edge that is not apparent in the observations. These discrepancies are discussed more in Section \ref{sect:61005_model_discussion}. We also note that the ``fringing'' seen in the back edge of the ring is just an artifact of finite particle number.

One of the most striking and significant aspects of the model is that the ring's intrinsic major axis appears as its minor axis when seen in projection on the sky. This is clear from Figure \ref{fig:grain_orbits}. High-eccentricity and large semimajor axis dust orbits, born together at the same pericenter, have their apoapses clustered toward one side of the star and pointed toward us as the observers. The result is a long ``fan'' that is not readily apparent when viewed at high inclination but ultimately creates many of the disk's morphological features. This result agrees with the analysis of \citet{lee2016} and is similar to a result from \citet{maness2009}, albeit produced by a different perturbation mechanism in the latter.

Examining the model's morphology more quantitatively, we find that it agrees better with observations in some regards than in others. Measuring the ring center in the self-subtracted model with the same method used for the images, we find it to be offset from the star by $1.7\pm0.7$ AU to the W and $0.5\pm0.3$ AU to the S. This is statistically consistent with the offset in the $H$-band image reported in Section \ref{sect:61005_morphology}, though slightly less offset to the W. The offset persists in the raw model ($1.5\pm0.7$ AU to W, $0.5\pm0.3$ AU to S), supporting the idea that it is an intrinsic disk feature and not an artifact of self-subtraction. Our 0.7 AU errors represent shifts in the ring center of 2 pixels, with smaller 1-pixel shifts (0.3 AU) along the minor axis.

One clear difference between model and data is that the W wing is $\sim$50\% ``shorter'' in the model than in the data when measured from the \editbf{``elbow'' previously described in Section \ref{sect:61005_morphology}}. The model wing starts at a radius of $\sim$62 AU from the star, and its brightness decreases to zero at $\sim$98 AU, while the observed (NIRC2 $H$) wing extends from $\sim$67 to $\sim$135 AU before reaching the background level in the data. Similarly, the model's E wing is also $\sim$40\% shorter than in the data (62--100 AU vs. 62--127 AU).  This is the most glaring discrepancy between our model and observations but may be more a result of our particular implementation of the model rather than a failure of the general planet-perturbed disk model. This is explored further in Section \ref{sect:61005_model_discussion}.

\begin{figure}[ht]
\centering
\includegraphics[width=\columnwidth]{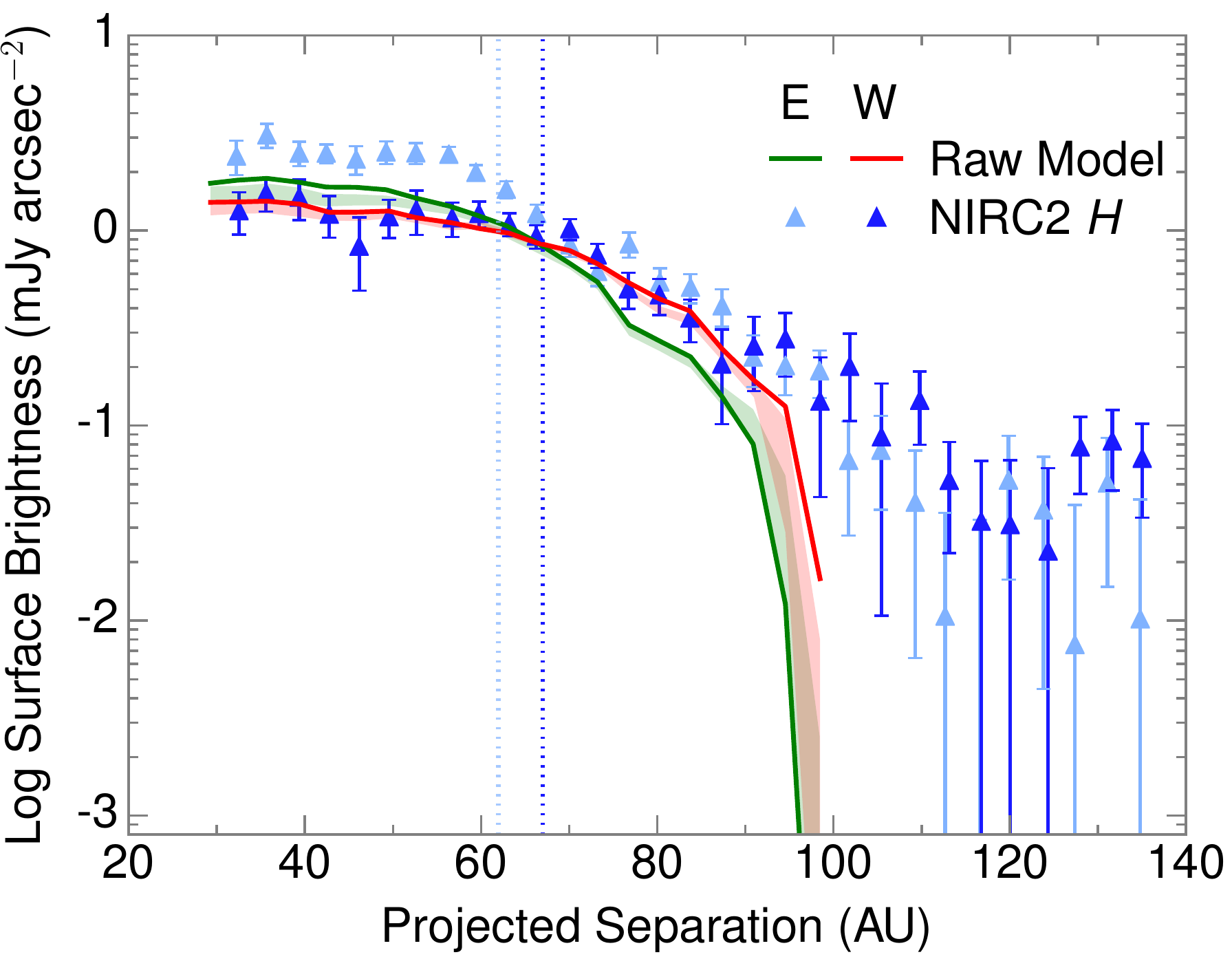}
\caption{Surface brightness radial profiles for the raw best-fit model (solid lines). Shading marks the 16th and 84th percentile brightness measurements from 1000 models drawn randomly from the MCMC walker chains. Self-subtraction-corrected $H$-band NIRC2 (blue) profiles are shown for comparison. The ``elbows'' at the junctions between ansae and wings in the observations are marked as light (E) and dark (W) dotted lines. \editbf{The raw model brightness, not biased by ADI self-subtraction, is roughly consistent with the data on the W side of the disk out to the end of the model wing at $\sim$100 AU, but underpredicts the E brightness by up to a factor of $\sim$2.}} 
\label{fig:sb_models}
\end{figure}

\editbf{In terms of surface brightness, the model agrees better with the data on the W side of the disk than on the E side.} This is demonstrated by Figure \ref{fig:sb_models},
in which we plot brightness profiles measured from
the raw best-fit model using the same aperture
method and positioning as for the NIRC2 data (but
not needing any self-subtraction correction for the
model). The shaded regions represent the 16th and 84th percentile brightness measurements among 1000
models drawn randomly from our MCMC walker chains,
and the $H$-band profiles are plotted again for
comparison. \editbf{The model agrees well with the data in the W throughout the ring, ansa, and wing. However, the model is typically 1.5--2 times fainter than the data in the E ring and ansa, with slightly better agreement in the inner part of the wing (we consider each wing to be everything exterior to the elbow, marked by a dashed line). This deficit in the east weakens the model ring's E $>$ W asymmetry (a factor of $\leq$1.3 difference) compared to observations (factor of 1.5--2.5).} Our fitted cubic
spline function $CS$ varies by
more than a factor of 10 across $\phi$,
indicating that parent
bodies have not fully differentially precessed
and are not in steady state. Together
with the modest rotation of the disk's major axis
away from our line of sight, this helps
to explain some but not all
of the E $>$ W asymmetry.

We can also explore how the values of specific parameters affect the model disk's morphology and brightness. We will mainly discuss parameters in terms of their best-fit values, as the main characteristics of the model vary little between the best-fit, 16\%, and 84\% likelihood parameter sets. For example, the median likelihood parameters produce a model that is nearly identical in appearance to the best-fit model and has a similar $\chi_\nu^2$ of 1.16. 

Our best-fit model has a parent body semimajor axis of $a=40.4$ AU, interior to the inner edge of the scattered-light ring in both model and data. This was paired with a best-fit value of $\beta=0.26$, resulting in dust particles pushed by radiation pressure to semimajor axes of 76--115 AU. Those two parameters are highly covariant, as a larger $\beta$ will increase the effects of radiation pressure on the dust and make up for a smaller $a$. To a lesser extent, we found both $a$ and $\beta$ to be degenerate with $e_{p0}$ and $e_f$. This is understandable, as changes in eccentricity will also move dust closer to or farther from the star. Together, these four parameters are primarily responsible for setting the true size of the ring and inner clearing (momentarily ignoring projection effects from inclination).

The best-fit values of $e_{p0}=0.08$ and $e_f=0.21$, 
together with $\beta=0.26$,
resulted in final dust eccentricities of 0.54--0.75.
Many of these high-eccentricity orbits also have large semimajor axes, and it is these dust
particles that fill in the ``fan'' that extends in front of the star. 
The secular perturbation theory we use states
that $e_{per} > e_f$ (the farther the planet is from the parent body,
the more eccentric it must be), so the implication of our model
is that the underlying planet is substantially eccentric.
Further broad constraints on the planet's mass and semimajor
axis are discussed in Section
\ref{sect:planet_constraints}.

Viewing geometry plays an important role. 
The $I_{per}=95\fdg7$ inclination of the planet orbit
to the sky plane and additional $I_0=4\fdg2$ 
mutual inclination between parent body orbits and the planet
are responsible for multiple features of the model disk's morphology.
The low opening angle of the ring requires $I_{per}$ to be close
to $90\arcdeg$ and the value is well constrained by detection of nearly the complete ring in the data. Alone, an $I_{per}$ a few degrees greater than $90\arcdeg$ is sufficient to produce a swept-back shape in the disk, even if the planet and parent bodies are coplanar (i.e., $I_0=0\arcdeg$). The effect of the inclination is to rotate the front fan so it extends several degrees south of the star when seen in projection, thus creating wings. In this particular system, however, we find that the wing shape better approximates the data when $I_{per}$ works in concert with a nonzero $I_0$. This fan is the dominant source of scattered light for the wings and dust apron, meaning that the parameters $a$, $\beta$, $e_{p0}$, and $e_f$ also play vital roles in creating those features. Combinations of those parameters that extend the fan farther from the star (e.g., higher $e_f$) will increase the radial size of the wings and apron, though this may distort other model features as a consequence.

The values for $\Omega$ and $\omega$ imply
that the planet's and disk's apastra are generally
pointed toward the observer. More precisely,
the initial parent body orbits' apastra 
(equivalently dust grains' apastra) are rotated several degrees to the
east (as viewed at the observed inclination). This rotation --- encapsulated
in the cubic spline function, which gives greater
weight to some parent body orbits over others --- is one of
the contributors to the brightness asymmetry,
as it is propagated to the dust orbits and augments the dust particles
with low scattering angles on the E side of the disk (thereby increasing
the number of forward-scattered photons) while depleting them on the W
side. However, this is a relatively weak effect in our model and does
not create enough asymmetry to match the data. As may be expected, we
found some degeneracy between various rotational angles, particularly
within the pairs of angles for planet and parent body.

\section{DISCUSSION} \label{sect:61005_discussion}

\subsection{Observations} \label{sect:61005_disc_obs}

Overall, our NIRC2 observations confirm the disk features reported in previous works, particularly those derived from the ADI $H$-band observations of \citet{buenzli2010}. Our $J$ and $K_p$ imaging results are similar to those in $H$. This continuity of disk characteristics from 1.2--2.3 $\mu$m implies that these different wavelengths are probing a single dust population comprising grains with scattering properties that are only weakly dependent on wavelength beyond an overall albedo trend accounting for the global blue color.

Here we discuss the results of our imaging but leave discussion of the swept-back morphology for the next section, as it is highly relevant to our modeling efforts.

\textit{Ring Geometry}: \editbf{the offset between the breaks in the brightness radial profiles is an interesting feature.} Both $H$ and $K_p$ show that the W profile's \editbf{break} is $\sim$10 AU farther from the star than the E \editbf{break} is. A total of 5--6 AU of that shift can be attributed to the stellocentric offset. However, that leaves 4--5 AU to explain. This is two or three resolution elements at these wavelengths, so this shift is significant. It may be a geometric viewing effect due to different lines of sight to the two ansae because the ring's major axis is not pointed directly at us. On the other hand, it could be a physical feature of the dust in the ring, such as a local overdensity in the E ansa. A single large collision between planetesimals or an enhanced collision rate between smaller bodies could theoretically produce greater quantities of dust in a specific location. This question deserves more attention in the future when larger telescopes and finer modeling can probe yet smaller scales.

\textit{Disk Color}: the disk's mean $H-K_p\approx-1.0$ mag color makes it distinctly blue compared to many other debris disks. This near-IR color is very similar to the mean [F606W]$-$[F110W]$=-1.2 \pm 0.3$ mag color presented by \citet{maness2009}. As those authors suggest, this implies that the disk's dust population contains a larger number of grains at increasingly small sizes and is dominated by $\sim$0.1--2 $\mu$m grains that scatter efficiently at optical/near-IR wavelengths. A similar argument was made for the AU Microscopii debris disk, with measured colors of $V-H<-1$ mag and $H-K_p\lesssim-0.5$ mag \citep{krist2005, augereau2006, fitzgerald2007_aumic, graham2007}. Our measurement for HD 61005 makes it even bluer in the near-IR than AU Mic, calling for particularly small grains or a composition that intrinsically produces blue scattering. Coupled with the strong forward-scattering suggested by the GPI data, this may present a particular challenge to model.

The disk's blue color is roughly constant with radial separation, suggesting that the small grain populations are the same in the ring and in the wings. This could be explained by a disk that is intrinsically homogeneous and well mixed radially. Such a composition may arise when small grains are produced in the ring by collisions between parent bodies and then are blown outward by radiation pressure onto more eccentric orbits. This scenario would be consistent with our perturbed disk model, which produces the scattered-light signal of both the ring and the wings from the same dust population.

\textit{Inner Disk}: the GPI data reveal the innermost regions of the HD 61005 disk. The ring appears to continue smoothly with increasing brightness from $>$40 AU in to the speckle-limited inner working angle of $\sim$9 AU projected separation. A peak in the scattered-light brightness at the smallest separations indicates that the ring is composed of primarily forward-scattering grains, a characteristic shared by many other resolved disks. Some bright clumps are visible in the ring but they are of low significance and may be the result of KLIP subsection positioning. The lack of stronger clumpy structure, such as that seen in the AU Mic disk \citep{fitzgerald2007_aumic, boccaletti2015a, wang2015_aumic}, suggests that the ring's dust distribution is relatively smooth. This may argue against a planetary body orbiting within the ring itself, where it might carve out gaps or push dust into resonance traps.

\textit{Polarization}: the $Q_r$ image shows additional light south of the ring compared to the total intensity image. As noted earlier, low-frequency features like this are often filtered out by algorithms like LOCI and KLIP. Thus, this polarized light may be coming from the smooth ``apron'' of dust seen extending south of the star in STIS imaging \citep{schneider2014}. We may not see the signal continuing farther south because, with a $\lesssim$10\% polarization fraction, the outer parts of the apron may be too faint for GPI to detect in polarized intensity.

The ring's polarization fraction is a few percent at projected separations of tens of AU and shows a trend of increasing with separation starting at $\sim$35 AU. A similar upward trend is seen in ACS observations of this disk and, among other examples, in recent GPI observations of the HD 111520 debris disk \citep{draper2016}. The similar polarization properties imply that these disks may contain grains with comparable attributes. HD 111520 is also highly inclined (nearly edge-on) and displays a ``needle'' morphology, possibly indicating another planet-perturbed system. More can be learned about grain size, shape, and composition from the polarization properties of the dust in the HD 61005 disk, but in-depth investigation of such a complex topic is outside the scope of this work.

\subsection{Model} \label{sect:61005_model_discussion}

Our best-fit model reproduces major features of resolved
scattered-light images of
HD 61005. There are three physical ingredients in the model: a planet
secularly perturbs planetesimals (parent bodies); collisions
among planetesimals produce small dust grains; radiation pressure
from the star perturbs dust grains onto highly eccentric
(but still bound) orbits. Notably, we do not require any interaction
between the disk's particles and ISM gas to 
reproduce the observed morphology. Here we discuss several aspects 
of the model, focusing on those that could be improved on
in future work.

\textit{Wing Length:} one discrepancy between the data and our
scattered-light models concerns the abbreviated lengths of the
model disk's wings. A contributing cause to this discrepancy
is that the high-S/N ring dominates $\chi^2$, while the S/N
of the wings decreases with distance from the ansae.
Less weight is therefore given to the outer parts of the wings
during the fitting process.

But there are also physical problems with the model that, if remedied,
could result in an improved fit. We model only a single population
of parent bodies that give birth to dust particles having only
a single size, and therefore only a single $\beta$,
at a single orbital phase (periastron). A smooth distribution of dust grains and $\beta$ values extending to
the radiation blowout limit would be more realistic \citep{strubbe2006}.
Radiation $\beta$ values larger than $\sim$0.3 would lengthen the wings
to $\sim$120 AU in projected separation and would also extend
the apron farther south of the star to better reproduce
the \textit{HST} data. Simulations of collisions (e.g., \citealt{lithwick2007, stark2009, nesvold2015}) and studies
of collisional cascades (e.g.,
\citealt{krivov2005, shannon2011, pan2012})
promise new modeling directions.

\textit{Ring Asymmetry:} as the radial surface brightness
profiles in Figure \ref{fig:sb_models} indicate,
our best-fit model underestimates the
brightness of the E front edge
of the disk and thus does not quite reproduce the observed factor-of-two
E $>$ W brightness asymmetry. 
If the lack of a detection of the E back side
of the ring is the result of a real deficit in brightness there,
rather than ADI over/self-subtraction,
then it could be linked to the faint W
front edge. The E back side may trace the periastra of a set
of dust orbits whose apastra trace the W front side.
Conversely, the W back side may trace the periastra of a second
set of dust orbits whose apastra trace the E front side.
To explain the E$>$W brightness asymmetry, there would need
(for some reason)
to be more dust grains in the first set of orbits than the second.
Our spline function that assigns different weights to orbits
having different precession angles $\phi$ --- thereby allowing
for particles that have not yet equilibrated secularly \citep{olofsson2016} --- can account
for some but not all of the brightness asymmetry.
Relaxing our assumption of a single fixed proper eccentricity $e_{p0}$
should help (i.e., allowing for a locus of orbits
in complex eccentricity and inclination
space that is not strictly circular; see Figure \ref{fig:parent_e_i}).
Other ingredients missing from our model that might
be relevant include light-scattering phase functions
that account for different grain sizes and/or compositions,
and multiple planets (e.g., \citealt{wyatt1999}).

\textit{Parent Body Ring, at NIR and Longer Wavelengths:} the various
model parameters listed in Table 2
indicate that parent bodies are distributed in an elliptical ring
extending from $\sim$30 AU (periastron)
to $\sim$70 AU (apastron). These findings connect
well with \citep{ricarte2013}, who infer that the bulk of
the disk's thermal millimeter-wave emission originates from
bodies located $\sim$60 AU from the star.
We also appear roughly consistent with \citet{steele2016},
who combine marginally resolved millimeter-wave images with the disk's SED to infer a dust belt of radius $\sim$60--70 AU. Recent 1.3 mm ALMA data presented by \citet{olofsson2016} also indicated parent bodies with semimajor axes of $\sim$66 AU. Notably, the disk's wings are not detected in the ALMA data despite sufficient angular resolution to do so, which is consistent with our model's distribution of parent bodies in the ring and only dust in the wings.

\textit{Edges and Spurs:} a feature that we find in our models but not in the data is a
second bright ``edge'' along the bottom of the apron. This appears to be
caused by many dust apoapses overlapping, not at the outer edge of the fan
but at its inner edge. These are the lower $e_d$, smaller $a_d$ orbits
that remain closer to the star but are still apsidally aligned. Dust particles slow down and bunch up near apoapse, creating local enhancements in optical depth and thus a bright edge. \editbf{In the real disk, multiple dust populations that are less apsidally aligned and have more varied apastra may smooth this feature and reduce its brightness below detection limits.}

As noted in Section
\ref{sect:model_results}, our raw models do not contain the radial
``spurs'' seen to extend outward from the ansae in the
ACS and STIS data \citep{maness2009, schneider2014}.
These spurs might be related to the ``double wing'' morphology
found by \citet{lee2016}, features that depend on extended
distributions of $\beta$ and apoapse distance
that our single-$\beta$ model lacks.

\begin{figure}[h]
\centering
\includegraphics[width=\columnwidth]{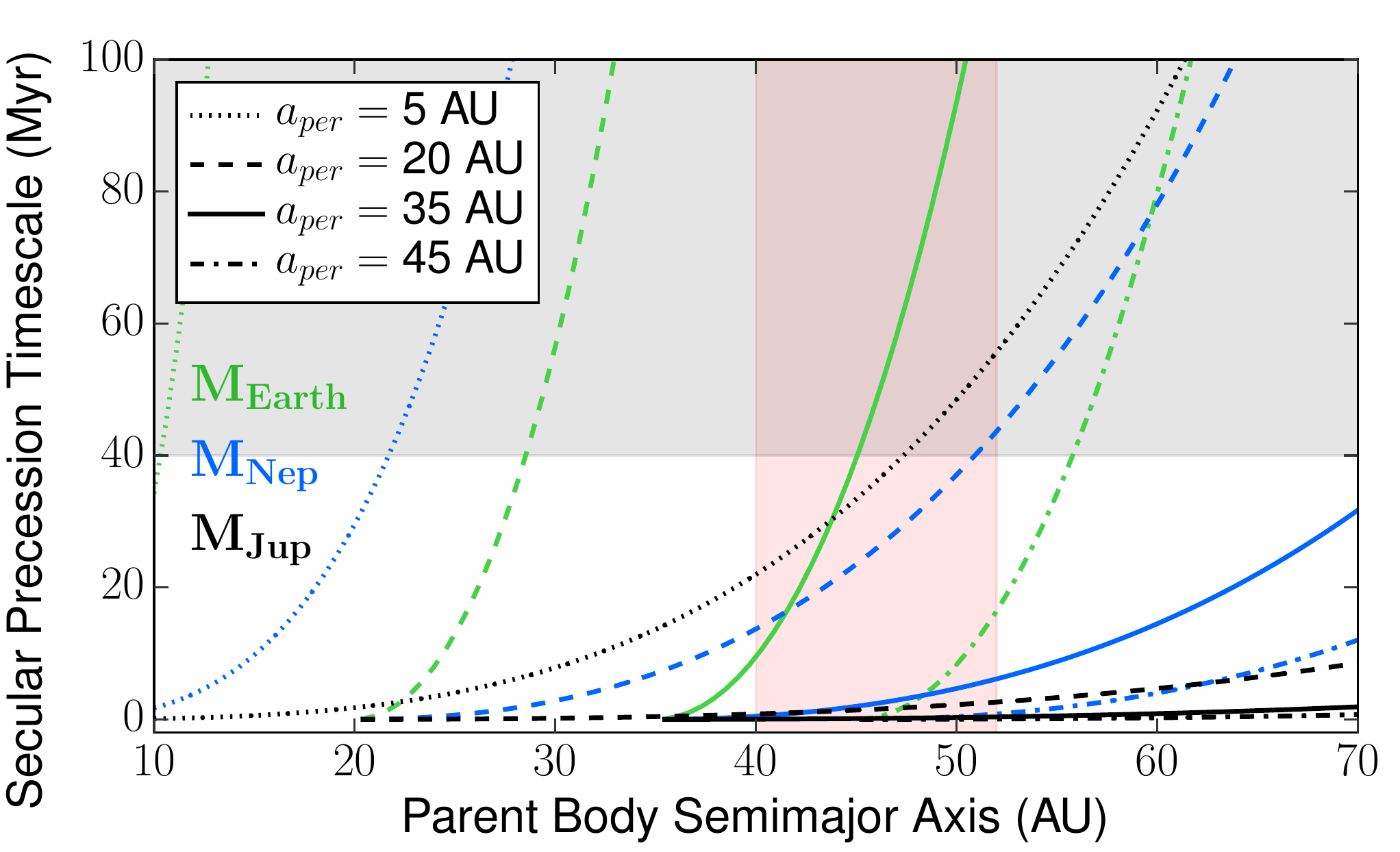}
\caption{Secular precession timescales for disk parent bodies at various semimajor axes $a$ as a function of planet mass and planet semimajor axis. The curves show that many mass--semimajor axis permutations will produce precession timescales shorter than the minimum estimated system age of 40 Myr (white region), which we consider sufficient to have shaped the disk's morphology. Line colors denote the planet mass, plotted for Earth (green), Neptune (blue), and Jupiter (black) masses. Line styles denote the planet semimajor axis. The red shaded region marks a range in parent body semimajor axis from the best-fit value to the 84th percentile value. For bodies at the best-fit $a$ of $\sim$40 AU, examples of possible disk-shaping planet masses and semimajor axes include $\mathrm{M}_\oplus$ at 35 AU, $M_\mathrm{{Nep}}$ at 20 AU, or $M_\mathrm{{Jup}}$ at 5 AU.}
\label{fig:tsec}
\end{figure}

\subsection{Planet Constraints} \label{sect:planet_constraints}

In the context of our secular perturbation model,
we most strongly constrain the planet's eccentricity.
According to Laplace--Lagrange secular theory, as $a$ approaches $a_{per}$, $e_f$
approaches $e_{per}$. Therefore, $e_f$ sets a lower limit for the planet's
eccentricity. Our fitted values of $e_f \sim 0.21$--0.27 indicate
a high planet eccentricity, not unlike those of giant planets discovered by radial velocity surveys (e.g., \citealt{zakamska2011} and references therein).

Other planet properties are only weakly constrained.
If we assume that the secular precession period of a parent body
is less than the system age --- estimated to be $\gtrsim$ 40 Myr --- there
are many combinations of planet mass and planet semi-major axis
that are allowed, as Figure \ref{fig:tsec} demonstrates.
The red shaded region in Figure \ref{fig:tsec} delimits parent body semimajor axes,
ranging from the MCMC best-fit value to the 84th percentile value.
Multiple curves cross the red shaded region below the 40 Myr mark,
demonstrating that there exist many mass--semimajor axis permutations
capable of shaping the disk within that timescale.
For example, a Neptune-mass planet with $a_{per}=20$ AU would require only 14 Myr to perturb parent bodies at the best-fit $a$ of $\sim$40 AU.
If the system were instead 100 Myr old, then the allowed
parameter space opens further, in particular to include smaller planet masses. \editbf{Our observations do not substantially reduce this parameter space, as they were primarily sensitive to planets more massive than $\sim$1.5 $M_\mathrm{{Jup}}$ with $a_{per}\gtrsim10$ AU on projected orbits that take them away from the disk brightness (Figure \ref{fig:contrast}), which our modeling indicates is not the preferred case.}

\subsection{Differentiating the Planet-perturbation and ISM-interaction Models}

The ISM can secularly perturb 
small dust grains bound to the host star
and produce a moth-like morphology in scattered light
\citep{maness2009}. The monodirectional flow of the ISM
across the disk induces a global disk eccentricity,
mimicking some of the effects of an eccentric perturbing
planet. This same ram pressure from the ISM should
affect parent bodies (having smaller area-to-mass
ratios) less; thus, in ISM-interaction models, there is no
reason to expect that-longer wavelength (e.g., millimeter-wave)
images tracing larger bodies
should exhibit any stellocentric offset.
By comparison, in models like ours involving an eccentric planet, the offset should decrease toward longer wavelengths but should remain nonzero.
So far, millimeter-wave images lack the resolution
to decide this issue \citep{ricarte2013, olofsson2016, steele2016}.

Of course, planetary and ISM perturbations
are not mutually exclusive. No single model has yet reproduced
all the features seen in HD 61005;
see section \ref{sect:61005_model_discussion}.
The fact that the star's proper motion points north \citep{vanLeeuwen2007}, while the disk's swept-back wings are directed to the south,
might be more than just a coincidence.

\section{CONCLUSIONS} \label{sect:conclusions}

The unusual morphological features observed in the HD 61005 debris disk over the past decade have made it a particularly interesting case study for physical mechanisms driving those features. In this work, we combined high-resolution near-IR imaging with multidimensional modeling to demonstrate that the observed morphology could be the result of secular perturbations from a yet undetected planet residing in the system.

The new $J$, $H$, and $K_p$ scattered-light images from Keck/NIRC2 that we presented offer the highest angular resolution view of the disk to date. We also presented GPI $H$-band total intensity and polarized intensity data that probe the system down to projected separations of $<$10 AU. Together, the data illustrate the characteristics of both the inner and outer disk. These characteristics include: 
\begin{enumerate}
    \item A dust ring with a sharp inner edge at $\sim$50 AU and a projected stellocentric offset of $\sim$2 AU. The ring's front edge is up to 2.5 times brighter E of the star than W, and is substantially brighter than the weakly detected back edge. The front edge's brightness also increases as projected separation decreases down to our inner working angle of $\sim$10 AU.
    \item Swept-back wings extending $\sim$65 AU in projection from the ring ansae and deflected south of the ring at an angle of $\sim$22$\arcdeg$ ($\sim$3$\arcdeg$ steeper in W than E). Unlike the ring, the two wings have similar brightnesses.
    \item Roughly uniform morphological features among our three near-IR wavebands but an $H-K_p$ color that is distinctly blue throughout the disk, suggesting a single dust population consisting of small grains that preferentially scatter shorter wavelengths.
\end{enumerate}

To explain the primary morphological features of the disk, we employed a model in which a planet on an eccentric orbit secularly perturbs a mutually inclined exterior ring of large parent bodies. Those bodies then spawn small dust particles on inclined, eccentric orbits that scatter starlight and produce the disk's near-IR surface brightness. We used an MCMC simulation to compare a large sample of these scattered-light models with our NIRC2 $H$-band data and estimate the most likely values for 17 model parameters.

The resulting best-fit and median likelihood models approximately reproduced the offset ring, swept-back wings, and brightness asymmetries of the data. They also display the apron of dust seen filling the space between the wings south of the star in space-based data. We accomplish this result without including any interaction between the disk and the surrounding ISM. Notably, key features, such as the clearly defined wings and apron, arise from a fan of dust extending toward the observer and composed of apsidally aligned dust orbits with high eccentricities and large semimajor axes that are viewed nearly edge-on.

Our highest-likelihood models indicate that the system consists of parent bodies with semimajor axes of $\sim$40--50 AU, proper eccentricities of $\sim$0.1, and mutual inclinations of $4\arcdeg$ relative to the planet's orbit (which itself is $\sim6\arcdeg$ from edge-on to the observer). These parent bodies give rise to forward-scattering ($g=0.6$) dust particles of $\beta\approx0.25$ that attain semimajor axes and total eccentricities of 76--115 AU and 0.54--0.75, respectively. We find both the parent body and dust populations to be located roughly consistently with locations predicted by long-wavelength observations and SED models.

This model only weakly constrains the theoretical planet's properties, though we require it to be apsidally aligned with the parent bodies and have a semimajor axis smaller than said bodies (i.e., no greater than $\sim$50 AU). It is also predicted to have at least a moderate eccentricity of $\sim$0.2. Based on the length of secular precession timescales needed to shape the disk within the system's age, we can jointly constrain planet mass and semimajor axis for a given parent body population, with a Neptune-mass planet at 20 AU being one possible solution.

While we have demonstrated that planet-driven perturbations may be responsible for the HD 61005 debris disk's morphology, work remains to assess whether other disks with resolved structures like ``needles'' and ``double-wings'' can be explained in the same way. This will be possible with the future application of enhanced models containing additional layers of sophistication to even more informative data from the current and coming generation of high-contrast imaging instruments.

\acknowledgments

{\bf Acknowledgements:} The authors wish to thank the anonymous referee for helpful suggestions that improved this manuscript. T.~E. was supported by a UCLA graduate research fellowship and in part by NASA Grants NNX14AJ80G, NNX15AC89G, and NSF AST-1518332. M.~P.~F. and G.~D. recognize the support of the NSF (AST-1413718) in their work on GPIES. Portions of this work were performed under the auspices of the U.S. Department of Energy by Lawrence Livermore National Laboratory under Contract DE-AC52-07NA27344. The results reported herein also benefited from collaborations and/or information exchange within NASA's Nexus for Exoplanet System Science (NExSS) research coordination network sponsored by NASA's Science Mission Directorate grant NNX15AD95G.
\par This work used computational and storage services associated with the Hoffman2 Shared Cluster provided by UCLA Institute for Digital Research and Education’s Research Technology Group. It also has made use of the NASA Exoplanet Archive, which is operated by the California Institute of Technology, under contract with the National Aeronautics and Space Administration under the Exoplanet Exploration Program.
\par Some of the data presented herein were obtained at the W.~M. Keck Observatory, which was made possible by the generous financial support of the W. M. Keck Foundation and is operated as a scientific partnership among the California Institute of Technology, the University of California, and the National Aeronautics and Space Administration. The authors wish to recognize and acknowledge the very significant cultural role and reverence that the summit of Mauna Kea has always had within the indigenous Hawaiian community. We are most fortunate to have the opportunity to conduct observations from this mountain.

\editbf{\software{Gemini Planet Imager Data Pipeline (\citealt{perrin2014_drp}, \url{http://ascl.net/1411.018}), pyklip (\citealt{wang2015_pyklip}, \url{http://ascl.net/1506.001}),
emcee (\citealt{foreman-mackey2013}, \url{http://ascl.net/1303.002})}.}

\facility{Keck:II (NIRC2), Gemini:South (GPI)}


\clearpage

\bibliographystyle{aasjournal}
\bibliography{disk_exop_refs}

\end{document}